\documentclass[final]{siamltex}
\usepackage{subfig, amsmath, amsfonts}
\usepackage{graphicx,color}

\newcommand{\bvec}[1]{\mathbf{#1}}

\newcommand{\mcV}{\mathcal{V}}

\newcommand{\Vstar}{V^{\ast}}
\newcommand{\Jstar}{J_{\ast}}
\newcommand{\tJstar}{\tilde{J}_{\ast}}

\newcommand{\xc}{\mathrm{xc}}
\newcommand{\vr}{\bvec{r}}

\newcommand{\vq}{\bvec{q}}

\newcommand{\ud}{\,\mathrm{d}}

\newcommand{\Vxc}{V_{\mathrm{xc}}}
\newcommand{\Kxc}{K_{\mathrm{xc}}}
\newcommand{\vopt}{V_{\mathrm{opt}}}

\newcommand{\Vion}{V_{\mathrm{ion}}}
\newcommand{\abs}[1]{\lvert#1\rvert}
\newcommand{\norm}[1]{\lVert#1\rVert}

\newcommand{\ie}{\textit{i.e.}{}}

\newcommand{\Or}{\mathcal{O}}
\newcommand{\mcF}{\mathcal{F}}

\newcommand{\lmin}{\lambda_{\min}}
\newcommand{\lmax}{\lambda_{\max}}

\title{Elliptic preconditioner for accelerating the self consistent field iteration in Kohn-Sham density functional theory}

\author{
Lin Lin \thanks{
Computational Research Division, Lawrence Berkeley National
Laboratory, Berkeley, CA 94720. Email: linlin@lbl.gov} 
\and Chao Yang \thanks{Computational Research Division, Lawrence Berkeley National
  Laboratory, Berkeley, CA 94720. Email: cyang@lbl.gov}
}

\begin{document}

\maketitle

\begin{abstract}
We discuss techniques for accelerating the self consistent field (SCF) iteration for solving the Kohn-Sham equations.  These techniques are all based on constructing approximations to the inverse of the Jacobian associated with a fixed point map satisfied by the total potential. They can be viewed as preconditioners for a fixed point iteration.  We point out different requirements for constructing preconditioners for insulating and metallic systems respectively, and discuss how to construct preconditioners to keep the convergence rate of the fixed point iteration independent of the size of the atomistic system.  We propose a new preconditioner that can treat insulating and metallic system in a unified way.  The new preconditioner, which we call an elliptic preconditioner, is constructed by solving an elliptic partial differential equation.  The elliptic preconditioner is shown to be more effective in accelerating the convergence of a fixed point iteration than the existing approaches for large inhomogeneous systems at low temperature.
\end{abstract}

\begin{keywords} 
  Kohn-Sham density functional theory, self consistent field iteration,
  fixed point iteration, elliptic preconditioner
\end{keywords}

\begin{AMS}
  65F08, 65J15, 65Z05
\end{AMS} \pagestyle{myheadings}
\thispagestyle{plain}
\markboth{L. LIN AND C. YANG}{ELLIPTIC PRECONDITIONER}

\section{Introduction}\label{sec:intro}

Electron structure calculations based on solving the Kohn-Sham
density functional theory (KSDFT)~\cite{HohenbergKohn1964,KohnSham1965} play an important role
in the analysis of electronic, structural and optical properties of
molecules, solids and other
nano structures.  The Kohn-Sham equations define a nonlinear 
eigenvalue problem
\begin{equation}
  \begin{split}
  &H[\rho]\psi_{i} = \varepsilon_{i} \psi_{i},\\
  &\rho(\vr) = \sum_{i} f_{i} \abs{\psi_{i}(\vr)}^2, \quad \int
  \psi^{*}_{i}(\vr) \psi_{j}(\vr) \ud \vr = \delta_{ij},
  \end{split}
  \label{eqn:KS}
\end{equation}
where $\varepsilon_{i}$ are the Kohn-Sham eigenvalues (or quasi-particle 
energies) and $\psi_{i}$ are called the Kohn-Sham wavefunctions or orbitals.  
These eigenfunctions define the electron density $\rho(\vr)$, which in
turn defines the Kohn-Sham Hamiltonian 
\begin{equation}
H[\rho] = -\frac12 \Delta + \mcV[\rho] + \Vion,
\label{eqn:ksham}
\end{equation}
where $\Delta$ is the Laplacian operator, $\mcV(\rho)$ is a nonlinear 
function of $\rho$, and $\Vion$ is a potential function that is independent 
of $\rho$.  The parameters $\{f_i\}$ that appear in the definition of $\rho$,
which are often referred to as the occupation number, are defined by
\begin{equation}\label{eqn:FermiDirac}
  f_i = \frac{1}{1 + \exp( \beta(\varepsilon_{i} - \mu))},
\end{equation}
where $\beta$ is proportional to the inverse of the temperature $T$ and
$\mu$ is called the chemical potential chosen to ensure that $f_i$'s satisfy
\begin{equation}\label{eqn:FDconstraint}
  \sum_i f_i = \int \rho(\vr) \ud \vr = N,
\end{equation}
for a system that contains $N$ electrons.  
The right hand side of (\ref{eqn:FermiDirac}) is known as
the Fermi-Dirac function evaluated at $\varepsilon_i$. 
When $\beta$ is sufficiently large, the Fermi-Dirac function behaves like a 
step function that drops from 1 to 0 at $\mu$ (which lies between 
$\varepsilon_{N}$ and $\varepsilon_{N+1}$). Spin degeneracy is omitted
here for simplicity.

In this paper, we assume the Kohn-Sham system
(\ref{eqn:KS}) is defined within the domain $\Omega=[0,L]^3$ with
periodic boundary conditions, and the number of electrons $N$ is
proportional to the volume of the domain.

Because the eigenvalue problem (\ref{eqn:KS}) is nonlinear, it is
often solved iteratively by a class of algorithms called {\em self-consistent
field iterations} (SCF).  We will show in the following that
the SCF iteration can be viewed as a fixed point iteration applied to a 
nonlinear system of equations defined in terms of the potential 
$\mathcal{V}$ that appears in~\eqref{eqn:ksham} or the charge density $\rho$.
The function
evaluation in each step of the SCF iteration is relatively expensive. 
Hence, it is desirable to reduce the total number of SCF iteration by 
accelerating its convergence.  Furthermore, we would like the convergence
rate to be independent of the size of the system.  In the past few decades, 
a number of acceleration schemes have been
proposed~\cite{Anderson1965,Pulay1980,Kerker1981,DederichsZeller1983,HoIhmJoannopoulos1982,KudinScuseriaCances2002,RaczkowskiCanningWang2001,AngladeGonze2008,MarksLuke2008}.
However, none of the existing methods provide a satisfactory solution
to the convergence issues to be examined in this paper, especially 
the issue of size dependency.

The purpose of the paper is twofold. First, we summarize a number of
ways to accelerate the SCF iteration.
Many of the schemes we discuss already exist in both the physics and 
the applied mathematics
literature~\cite{KresseFurthmuller1996,KresseFurthmuller1996a,AngladeGonze2008,
Annett1995,YangGaoMeza2009,FangSaad2009,RohwedderSchneider2011,WalkerNi2009}.
We analyze the convergence properties of these acceleration schemes.
In our analysis, we assume a good starting guess to the charge density 
or the potential is available. Such a starting guess is generally not 
difficult to obtain in real applications.
As a result, the convergence of the SCF iteration
can be analyzed through the properties of the Jacobian operator 
associated with the nonlinear map defined in terms of the potential or
the density.
Acceleration schemes can be developed by constructing approximations
to the Jacobian or its inverse. These acceleration schemes can also
be viewed as preconditioning techniques for solving a system of nonlinear 
equations.

%

It turns out that the SCF iteration exhibits quite different 
convergence behavior for insulating and metallic
systems~\cite{GhosezGonzeGodby1997,PickCohenMartin1970}.
These two types of systems are distinguished by the gap between 
$\varepsilon_{N}$ and $\varepsilon_{N+1}$ as the number of electrons
$N$, or equivalently the system size increases to infinity. For
insulating systems,
\begin{equation}
  \lim_{N\to \infty} E_{g} > 0,
  \label{eqn:finitegap}
\end{equation}
where $E_g = \varepsilon_{N+1} - \varepsilon_{N}$, whereas for 
metallic systems, $\lim_{N\to \infty} E_{g} = 0$.
Different accelerating (or preconditioning) techniques are
required for insulating and metallic systems.  

The second purpose of this paper is to propose a new framework 
for constructing a preconditioner for accelerating the SCF iteration. 
The preconditioner constructed under this framework, 
which we call the elliptic preconditioner, 
provides a unified treatment of insulating and metallic systems.  
It is effective for complex materials that contain both an
insulating and a metallic component. 
This type of system is considered to be
difficult~\cite{RaczkowskiCanningWang2001} for a standard Kohn-Sham
solver, especially when the $\beta$ parameter in~\eqref{eqn:FermiDirac} 
is relatively large (or the temperature is low).

The paper is organized as follows.  In section~\ref{sec:fixedpt}, we
introduce the fixed point iteration for solving the Kohn-Sham problem, 
and the simple mixing method as the simplest acceleration method.  
More advanced preconditioning techniques are discussed in
section~\ref{sec:advance}. In section~\ref{sec:convergence}, we discuss the
convergence behavior of the acceleration methods for increasing
system sizes.  Based on these discussions, 
a new preconditioner called the elliptic preconditioner is presented 
in section~\ref{sec:elliptic}.  The performance
of the elliptic preconditioner is compared to existing techniques for
one dimensional model problems and a realistic three dimensional problem in
section~\ref{sec:numerical}.  We conclude and discuss future work 
in section~\ref{sec:conclusion}.

In this paper, the Kohn-Sham orbitals $\{\psi_{i}\}$ are assumed to be
in $H^{1}(\Omega)$.  In practical calculations, they are discretized in
a finite dimensional space such as the space spanned by a set of
planewaves.  As a result, each operator corresponds to a finite dimensional 
matrix.  
Our discussion in this paper is not 
restricted to any specific type of discretization of the Kohn-Sham 
orbitals.
To simplify our discussion, we will not distinguish operators defined on
$H^{1}(\Omega)$ from the corresponding matrices obtained from
discretization unless otherwise noted.  This applies to both
differential and integral operators.  
Neither will we distinguish integral operators from their
kernels.  For example, we may simply denote 
$f(\vr)=A[g](\vr)\equiv \int A(\vr,\vr') g(\vr')\ud \vr'$ by $f=Ag$ and
represent the operator $A$ by $A(\vr,\vr')$.

\section{Fixed point iteration and simple mixing}\label{sec:fixedpt}
It follows from the spectral theory that the charge density 
$\rho$ defined in (\ref{eqn:KS}) can be written as 
\begin{equation}
	\rho(\vr) = \left[I+e^{\beta(H[\rho]-\mu I)}\right]^{-1}(\vr,\vr).
\label{eqn:rhofix}
\end{equation}
Here $[\cdot](\vr,\vr)$ denotes the diagonal elements of a
matrix, and $I$ is an identity operator. That is, $\rho(\vr)$ is the
diagonal part of the Fermi-Dirac function evaluated at the Kohn-Sham
Hamiltonian. The right-hand side of (\ref{eqn:rhofix}) defines a fixed
point map from $\rho$ to itself.

A similar fixed point map is also defined implicitly in terms of the 
potential $V=\mcV[\rho](\vr)$, where
\begin{equation}
  \mcV[\rho](\vr) = \int 
	\frac{\rho(\vr')}{|\vr-\vr'|}\ud\vr'  + V_{\xc}[\rho](\vr),
  \label{eqn:Vform}
\end{equation}
where the first term corresponds to the electron-electron repulsion,
and $V_{\xc}[\rho]$ is a nonlinear functional of $\rho$ and is
known as the {\em exchange-correlation} potential that accounts for
many-body effects of the electrons.    
The choice of $V_{\xc}$ is not unique. A number of expressions 
are available in the physics
literature~\cite{CeperleyAlder1980,PerdewZunger1981,Becke1988,LeeYangParr1988,PerdewBurkeErnzerhof1996}.
However, for the purpose of this paper, we do not need to be concerned
with the explicit form of $V_{\xc}$.  It should be noted that
$V_{\xc}$ is often much smaller than the electron-electron repulsion
term in magnitude. If the Dirac exchange \cite{dirac} is used,
$V_{\xc}$ often contains a term proportional to $\rho^{1/3}$.

It follows from~\eqref{eqn:rhofix} and~\eqref{eqn:Vform} that $\rho$ is
implicitly a function of the potential $V$, which we will
denote by $\rho = F(V)$.  The analysis we present below and the
acceleration strategies we propose are applicable to both the density
fixed point map (\ref{eqn:rhofix}) and the potential fixed point map
\begin{equation}
  V = \mcV[F(V)].
  \label{eqn:Vfix}
\end{equation}
Without loss of generality, we will focus on the potential fixed point
map (\ref{eqn:Vfix}) in the rest of the paper. We remark that 
evaluating $\mcV[\rho]$ in~\eqref{eqn:Vform} requires solving a Poisson 
equation, but the computation of $F(V)$ requires either diagonalizing 
the Hamiltonian $H[\rho]$ or 
approximating the Fermi-Dirac function of $H[\rho]$ directly~\cite{Goedecker1999}.  
Therefore, computing $F(V)$ is much more costly than computing $\mcV[\rho]$.

The simplest method for seeking the solution of~\eqref{eqn:Vfix}
is the fixed point iteration.  In such an iteration, we start from 
some input potential $V_1$, and iterate the following equation
\begin{equation}
V_{k+1} = \mcV\left[ F(V_k) \right],
\label{eqn:fixiter}
\end{equation}
until (hopefully) the difference between $V_{k+1}$ and $V_k$ is sufficiently
small.

When $V_k$ is sufficiently close to the fixed point solution $\Vstar$,
we may analyze the convergence of the fixed point iteration~\eqref{eqn:fixiter}
by linearizing the function $\mcV [ F( \cdot )]$ defined in (\ref{eqn:Vfix})
at $\Vstar$.

If we define $\delta V_k = V_k - \Vstar$, subtracting $\Vstar$ from
both sides of (\ref{eqn:fixiter}) and approximating $\mcV [ F(V_k)]$ by
its first-order Taylor expansion at $\Vstar$ yields
\begin{equation}
  \delta V_{k+1}
  \approx
  \frac{\partial \mcV}{\partial V} \biggl|_{V=\Vstar} \delta V_k,
  \label{eqn:fixedpointerror}
\end{equation}
where $(\partial \mcV / \partial V)|_{V=\Vstar}$ is the Jacobian of 
$\mcV [F(V)]$ with respect to $V$ evaluated at $\Vstar$.

It follows from the chain rule that
\[
\frac{\partial \mcV}{\partial V} = \frac{\partial \mcV}{\partial \rho} \frac{\partial F}{\partial V}. 
\]
Taking the functional derivative of $\mcV[F(V)]$ given in~\eqref{eqn:Vform} 
with respect to $\rho(\vr)$ yields
\begin{equation}
  \frac{\partial \mcV}{\partial \rho}(\vr,\vr') = \frac{1}{|\vr-\vr'|} +
  \frac{\partial \Vxc}{\partial \rho}(\vr,\vr'),
  \label{eqn:dvdr}
\end{equation}
where the terms in Eq.~\eqref{eqn:dvdr} represent kernels of
integral operators evaluated at $\vr$ and $\vr'$. The first term on the
right hand side of (\ref{eqn:dvdr}) is the Coulomb kernel.  It will be
denoted by $v_{c}(\vr,\vr')$ below. The second term is the functional
derivative of the exchange-correction potential with respect to $\rho$.
It is often denoted by $K_{\xc}(\vr,\vr')$ and is a Hermitian matrix.

In the physics literature, the functional derivative of $F$ with respect 
to $V$ is often referred to as the independent particle polarizability 
matrix, and denoted by $\chi\left( \vr,\vr' \right)$.
At zero temperature, $\chi\left( \vr,\vr' \right)$ is given by the
Adler-Wiser formula~\cite{Adler1962,Wiser1963}
\begin{equation}
  \chi(\vr,\vr') = 2\sum_{n=1}^{N}\sum_{m=N+1}^{\infty}
  \frac{\psi_{n}(\vr)\psi_{m}^{*}(\vr)\psi_{n}^{*}(\vr')\psi_{m}(\vr')}{\varepsilon_{n}-\varepsilon_{m}},
  \label{eqn:alderwiser}
\end{equation}
where $(\varepsilon_i,\psi_i)$, $i = 1,2,...$, are the eigenpairs defined in
(\ref{eqn:KS}).  Note that $\chi$ is a Hermitian matrix, and
is negative semidefinite since 
$\varepsilon_n \leq \varepsilon_m$.

Eq.~\eqref{eqn:fixedpointerror} can be iterated recursively to yield
\begin{equation}
  \delta V_{k+1} \approx \left( \frac{\partial \mcV}{\partial \rho}\chi
  \right)^{k}
  \delta V_{1}.
  \label{}
\end{equation}
Therefore, a necessary condition that guarantees the convergence of
the fixed point iteration is 
\[
  \sigma\left( \frac{\partial \mcV}{\partial \rho} \chi \right)< 1,
\]
where $\sigma(A)$ is the spectral radius of the operator (or matrix) $A$.  

Unfortunately this condition is generally not satisfied as we will show
later.  However, a simple modification of the fixed point iteration 
can be made to overcome potential convergence failure as long as 
$\sigma \left( \frac{\partial \mcV}{\partial \rho}\chi \right)$ is bounded.

The modification takes the form
\begin{equation}
  V_{k+1}=V_{k}-\alpha\left(V_{k}-\mcV\left[ F(V_{k}) \right]\right),
  \label{eqn:simplemixingupdate}
\end{equation}
where $\alpha$ is a scalar parameter.  The updating formula given above
is often referred to as simple mixing.  When $V_j$ is sufficiently close
to $\Vstar$, the error propagation of simple mixing scheme is 
\begin{equation}
  \delta V_{k+1} \approx \delta V_{k} - \alpha\left( I- \frac{\partial
  \mcV}{\partial \rho}\chi\right)\delta V_{k}.
  \label{eqn:errorsimplemixing}
\end{equation}
Notice that $I-(\partial \mcV/\partial \rho)\chi$ is simply the 
Jacobian of the residual function $V - \mcV[F(V)]$ with respect to $V$.
We will denote this Jacobian by $J$. Its value at $\Vstar$ will be denoted
by $\Jstar$. In the physics literature, this Jacobian is often
referred to as a {\em dielectric} operator~\cite{Adler1962,Wiser1963},
and denoted by $\varepsilon$. Furthermore, when
$\frac{\partial \mcV}{\partial \rho}$ is positive definite, 

$\varepsilon$ only has real eigenvalues because it can be symmetrized
through a similarity transformation 
\[
\widetilde{\varepsilon} = \left(\frac{\partial
\mathcal{V}}{\partial \rho}\right)^{-1/2} \varepsilon
\left(\frac{\partial \mathcal{V}}{\partial \rho}\right)^{1/2} = I -
\left(\frac{\partial \mathcal{V}}{\partial \rho}\right)^{1/2} \chi
\left(\frac{\partial \mathcal{V}}{\partial \rho}\right)^{1/2},
\]
where the symmetrized dielectric operator $\widetilde{\varepsilon}$ is Hermitian and
has real eigenvalues.
We remark that the assumption that $\frac{\partial \mcV}{\partial \rho}$ is positive
definite may not always hold, especially when the material 
contains low electron density regions in which the 
exchange-correlation kernel $K_{\xc}$ contains large negative entries. 
However, because in general the product of $K_{\xc}$ and $\chi$ is 
much smaller in magnitude than $v_c \chi$, it is reasonable to
expect that the eigenvalues of $\varepsilon$ are close to those 
of $I - v_c \chi$, which are real.  This type of approximation is also 
used in section~\ref{sec:elliptic} where we discuss how to construct
an effective preconditioner for accelerating the fixed point iteration.


It follows from (\ref{eqn:errorsimplemixing}) that simple mixing will
lead to convergence if 
\begin{equation}
  \sigma\left( I-\alpha \Jstar \right) < 1.
  \label{eqn:alfabond}
\end{equation}
If $\lambda(\Jstar)$ is an eigenvalue of $\Jstar$, then the condition given 
in~\eqref{eqn:alfabond} implies that
\begin{equation}
  \abs{1-\alpha\lambda(\Jstar)} <  1.
\end{equation}
Consequently, $\lambda(\Jstar)$ must satisfy
\begin{equation}
  0< \alpha < \frac{2}{\lambda(\Jstar)}.
  \label{eqn:alfabond1}
\end{equation}
Note that (\ref{eqn:alfabond1}) is only meaningful when $\lambda(\Jstar) > 0$
holds. Therefore, $\lambda(\Jstar)>0$ is often referred to as the 
{\em stability condition} of a
material~\cite{BauernschmittAhlrichs1996,LuE2010,LuE2011}.  
Furthermore, when $\lambda(\Jstar)$ is bounded, it is always possible to 
find a parameter $\alpha$ to ensure the convergence of the modified fixed point iteration even though the convergence may be slow. 

We should comment that the stability condition $\lambda(\Jstar) > 0$
holds in most cases because $\Kxc \chi$ is typically much smaller 
in magnitude than $v_{c} \chi$. Note that $v_c$ is positive definite, 
and $\chi$ is negative semidefinite.  When the stability condition fails,
phase transition may occur, such as the transition from uniform electron
gas to Wigner crystals in the presence of low electron
density~\cite{Wigner1934}. Such a case is beyond the scope of the current study.
Nonetheless, $\lambda(J_{*})$ can become very large in practice even
when the stability condition holds,
especially for metallic systems of large sizes, as we will show in
section~\ref{sec:convergence}. A large $\lambda(J_{*})$ requires
$\alpha$ to be set to a small value to ensure convergence.  Even though
convergence can be achieved, it may be extremely slow.


\section{Preconditioned fixed point iteration and quasi-Newton acceleration}\label{sec:advance}

The simple mixing scheme selects $\alpha$ as a scalar in (\ref{eqn:simplemixingupdate}).
If we replace the scalar $\alpha$  by
the inverse of the Jacobian matrix of the function
$V - \mcV[F(V)]$ evaluated at $V_k$, we obtain a Newton's update of $V$.  
When $V_k$ is in the region where the linear approximation given by 
(\ref{eqn:fixedpointerror}) is sufficiently accurate, Newton's method
converges quadratically to the solution of (\ref{eqn:Vfix}).

\subsection{Jacobian-free Krylov Newton}
The difficulty with applying Newton's method directly is that 
the Jacobian matrix
$J_k = I - \frac{\partial \mcV[F(V)]}{\partial V}|_{V = V_k}$ 
or its inverse cannot be easily evaluated. However, we may apply a 
Jacobian-free Krylov Newton technique~\cite{KnollKeyes2004} to obtain Newton's update 

\[
\Delta_k = J_k^{-1}r_k, \ \ \mbox{where} \ \
r_k = V_{k}-\mcV\left[ F(V_{k}) \right],
\]
by solving the linear system
\begin{equation}
J_k \Delta_k = r_k
\label{eq:corrV}
\end{equation}
iteratively using, for example, the GMRES
algorithm~\cite{SaadSchultz1986}. The matrix vector
multiplication of the form $y \leftarrow J_k x$, which is required 
in each GMRES iteration, can be approximated by finite difference
\[
y \approx x - \frac{\mcV[F(V_k + \epsilon x)]-\mcV[F(V_k)]}{\epsilon},
\]
for an appropriately chosen scalar $\epsilon$.

The finite difference calculation requires one additional function 
evaluation of $\mcV[F(V_k + \epsilon x)]$ per GMRES step. Therefore,
even though Newton's method may exhibit quadratic convergence,
each Newton iteration may be expensive if the number of GMRES steps 
required to solve the correction equation (\ref{eq:corrV}) is large.
The convergence rate of the GMRES method for solving the linear
system~\eqref{eq:corrV} is known to satisfy~\cite{LiesenTichy2004}
\begin{equation}
  \norm{\Delta_{k}^{n}-\Delta_{k}} \le C \left(
  \frac{\sqrt{\kappa(J_{k})}-1}{{\sqrt{\kappa(J_{k})}+1}} \right)^{n}
  \norm{\Delta_{k}^{0}-\Delta_{k}},
  \label{eqn:convGMRES}
\end{equation}
where
$\kappa(J_{k})=\frac{\lambda_{\max}(J_{k})}{\lambda_{\min}(J_{k})}$ is
the condition number of $J_{k}$, and $\Delta_{k}^{n}$ is the
approximation of $\Delta_{k}$ at the $n$th step of the GMRES iteration.
As we will show in Section~\ref{sec:convergence}, the condition number
$\kappa(J_{k})$ can grow rapidly with respect to the size of the system,
especially for metallic systems.  Therefore the number of
iterations required by an iterative solver also grows with respect to 
the size of the system unless preconditioning strategies are employed.

\subsection{Broyden's and Anderson's method} \label{sec:broyden}

An alternative to Newton's method for solving (\ref{eqn:Vfix}) 
is a quasi-Newton method that replaces $J_k^{-1}$ with an approximate
Jacobian inverse $C_k$ that is easy to compute and apply.  In such
a method, the updating strategy becomes
\begin{equation}
V_{k+1} = V_k - C_k\left(V_k - \mcV[F(V_k)] \right).
  \label{eqn:quasiNewtonUpdate}
\end{equation}
The simple
mixing scheme discussed in the previous section can be viewed as a 
quasi-Newton method in which $C_k$ is set to $\alpha I$
(or equivalently as a nonlinear version of the Richardson's iteration).
More sophisticated quasi-Newton updating schemes can be devised by using
Broyden's techniques~\cite{Johnson1988} to construct better approximations to
$J_k$ or $J_k^{-1}$. In Broyden's second method, $C_k$
is obtained by performing a sequence of low-rank modifications 
to some initial approximation $C_0$ of the Jacobian inverse using 
a recursive formula~\cite{FangSaad2009,MarksLuke2008} 
derived from the following 
constrained optimization problem
\begin{align}
\min_{C}  \hspace{.5in} &\frac{1}{2}||C - C_{k-1}||^2_F \nonumber \\
\text{s.t. } \hspace{.5in} & S_k = CY_k, \label{bmin2}
\end{align}
where $C_{k-1}$ is the approximation to the Jacobian constructed
in the $(k-1)$th Broyden iteration. The matrices $S_k$ and $Y_k$ above 
are defined as 
\begin{equation}
S_k = ( s_k, s_{k-1}, \cdots , s_{k-\ell} ), \ \ 
Y_k = ( y_k, y_{k-1}, \cdots , y_{k-\ell} ), \label{SYk}
\end{equation}
where $s_j$ and $y_j$ are defined by $s_j = V_j - V_{j-1}$ and
$y_j = r_j - r_{j-1}$ respectively.

It is easy to show that the solution to (\ref{bmin2}) is
\begin{equation}
C_k = C_{k-1} + (S_k-C_{k-1}Y_k)Y_k^{\dagger},
\label{bupd2}
\end{equation}
where $Y_k^{\dagger}$ denotes the pseudo-inverse of $Y_k$, i.e.,
$Y_k^{\dagger} = (Y_k^TY_k)^{-1}Y_k^T$. We remark that in
practice $Y_{k}^{\dagger}$ is not constructed explicitly since we only 
need to apply $Y_{k}^{\dagger}$ to a residual vector $r_{k}$.  
This operation can be carried out by solving a linear least squares 
problem with appropriate regularization (e.g., through a 
truncated singular value decomposition).

A variant of Broyden's method is Anderson's
method~\cite{Anderson1965} in which
$C_{k-1}$ is fixed to an initial approximation $C_0$ at each 
iteration. It follows from Eq.~\eqref{eqn:quasiNewtonUpdate} that 
Anderson's method updates the potential as
\begin{equation}
  V_{k+1} = V_{k} - C_{0}(I-Y_{k}Y_{k}^{\dagger}) r_k -
  S_{k}Y_{k}^{\dagger} r_k,
  \label{eqn:Anderson}
\end{equation}
In particular, if $C_0$ is set to $\alpha I$, we obtain Anderson's
method 
\[
V_{k+1} = V_{k} - \alpha (I-Y_{k}Y_{k}^{\dagger}) r_k -
  S_{k}Y_{k}^{\dagger} r_k.
\]
commonly used in KSDFT solvers.

\subsection{Pulay's method}

An alternative way to derive Broyden's method is through a technique
called Direct Inversion of Iterative Subspace (DIIS). The technique is
originally developed by Pulay for accelerating a Hartree-Fock calculation 
~\cite{Pulay1980}. Hence it is often referred to as the {\em Pulay mixing} 
in the condensed matter physics community. The motivation of Pulay's method
is to minimize the difference between $V$ and $\mcV[F(V)]$ within 
a subspace $\mathcal{S}$ that contains previous approximations to $V$. 
In Pulay's original work~\cite{Pulay1980}, the optimal approximation to $V$ from 
$\mathcal{S}$ is expressed as $V_{\mathrm{opt}} = \sum_{j=k-\ell-1}^k \alpha_j V_j$,
where $V_j$ ($j = k-\ell-1,  ..., k$) are previous approximations to $V$, and
the coefficients $\alpha_j$ chosen to satisfy the constraint 
$\sum_{j=k-\ell-1}^k \alpha_j = 1$.  

When $V_j$'s are all sufficiently close to the solution of (\ref{eqn:Vfix}),
$\mcV[F(\alpha_jV_j)] \approx \alpha_j \mcV[F(V_j)]$ holds. Hence we may 
obtain $\alpha_j$ (and consequently $\vopt$) by solving the following quadratic 
program
\begin{equation}
\begin{array}{cc}
\min_{\{\alpha_j\}} & \| \sum_{j=k-\ell-1}^k \alpha_j r_j \|_2^2 \\
\mbox{s.t.}         &  \sum_{j=k-\ell-1}^k \alpha_j = 1,
\end{array}
\label{eqn:pulaymin}
\end{equation}
where $r_j = V_j - \mcV[F(V_j)]$.

Note that (\ref{eqn:pulaymin}) can be reformulated as an
unconstrained
minimization problem if $\vopt$ is required to take the form
$\vopt = V_k + \sum_{j=k-\ell}^k \beta_j (V_j - V_{j-1})$, where
$\beta_{j}$ can be any unconstrained real number.  Again, if we assume
$\mcV[F(V)]$ is approximately linear at $V_j$ and let 
$b = (\beta_{k-\ell},...,\beta_k)^T$, minimizing
$\| \vopt - \mcV[F(\vopt)] \|$ with respect to $\{\beta_j\}$ 
yields $b = -Y_k^{\dagger} r_k$, where $Y_k$ is the same as that defined in 
(\ref{SYk}).

In \cite{KresseFurthmuller1996a,KresseFurthmuller1996}, Pulay's method for updating $V$ is defined as
\begin{equation}
V_{k+1} = \vopt - C_0 (\vopt - \mcV[F(\vopt)]),
\label{eqn:Pulay}
\end{equation}
where $C_0$ is an initial approximation to the inverse of the Jacobian
(at the solution).  Substituting $\vopt = V_k - S_k Y_k^{\dagger} r_k$
into (\ref{eqn:Pulay}) yields exactly Anderson's updating 
formula~\eqref{eqn:Anderson}. 

\subsection{Preconditioned fixed point iteration}
If $V_{\ast}$ is the solution to (\ref{eqn:Vfix}), then subtracting it
from both sides of the quasi-Newton updating formula
\[
V_{k+1} = V_k - C_k\left(V_k - \mcV[F(V_k)] \right)
\]
yields
\begin{equation}
\delta V_{k+1} \approx \delta V_k - C_k J_{\ast} \delta V_k
= (I - C_k J_{\ast}) \delta V_k,
\label{eqn:dvupdate}
\end{equation}
where $J_{\ast}$ is the Jacobian of the function $V - \mcV[F(V)]$ at 
$V_{\ast}$ and $C_k$ is the approximation to $J_{\ast}^{-1}$ constructed
at the $k$th step.  If $C_k$ is a constant matrix $C$ for all $k$, we can
rewrite (\ref{eqn:dvupdate}) as
\begin{equation}
\delta V_{k+1} = (I-C J_{\ast})^k \delta V_1.
\label{eqn:pfixit}
\end{equation}

Ideally, we would like to choose $C$ to be $J_{\ast}^{-1}$ to
minimize the error in the linear regime.  However,
this is generally not possible (since we do not know $V_{\ast}$). 
However, if $C$ is sufficiently close to $J_{\ast}^{-1}$, we may view $C$ as a 
preconditioner for a preconditioned fixed point iteration defined by 
(\ref{eqn:pfixit}).

A desirable property for $C$ is that $|\sigma(I-CJ_{\ast})|<1$ or 
\begin{equation}
0 < \sigma(CJ_{\ast}) < 2.
\label{eqn:eigbound}
\end{equation}
Because $J_{\ast} = I - (\partial \mcV / \partial \rho) \chi$, we may 
construct $C$ by seeking approximations to $\partial \mcV / \partial \rho$
and $\chi$ first and inverting the approximate Jacobian in (\ref{eqn:dvupdate}).
This is the approach taken by Ho, Ihm and Joannopoulos
in~\cite{HoIhmJoannopoulos1982}, which is
sometimes known as the HIJ approach. The HIJ approach
approximates the matrix $\chi$ by using Alder-Wiser formula given in
Eq.~\eqref{eqn:alderwiser} which requires computing all eigenvalues and 
eigenvectors associated with the Kohn-Sham Hamiltonian defined at $V_k$.  
The resulting computational cost for constructing $\chi$ alone is $\Or(N^4)$ 
due to the explicit construction of each pair of $\psi_n(\vr)$ and 
$\psi_m(\vr)$ for $n = 1,2, ..., N$ and $m = N+1, N+2,...$.
Such a preconditioning strategy is not practical for large problems.

An alternative to the HIJ approach is to use the ``extrapolar'' method
proposed in ~\cite{AngladeGonze2008}.  This method replaces $\psi_{m}(\vr)$
by planewaves for large $m$.  As a result, the number of $\psi_{m}$'s
that needs to be computed is reduced.  However, such a reduction
does not lead to a reduction in the computational complexity of constructing
$\chi$, which still scales as $\Or(N^4)$.  Therefore, the preconditioning 
strategy will become increasingly more expensive as the system
size increases.

A more efficient preconditioner that works well for simple metallic systems
is the Kerker preconditioner~\cite{Kerker1981}.  The potential updating scheme 
associated with this preconditioner is often known as the Kerker mixing 
scheme.  The construction of the Kerker preconditioner is based on 
the observation that the Coulomb operator $v_c$ can
be diagonalized by the Fourier basis (planewaves), and
the eigenvalues of the Coulomb operator are
$4\pi/q^2$, where $q = |\vq|$ is the magnitude of a sampled wave vector
associated with the Fourier basis function of the form $e^{i \vq \cdot
\vr}$.  Furthermore, for simple metals the polarizability operator
$\chi$ can be approximately diagonalized by the Fourier basis.  The
eigenvalues of $\chi$ are bounded from below and above.  Therefore,
if we omit the contribution from $\Kxc$, the
eigenvalues of $\Jstar$ are $1 + 4\pi \gamma/q^2 = (q^2 + 4\pi
\gamma)/q^2$ for some constant $\gamma>0$ which is related to the
Thomas-Fermi screening length~\cite{Ziman1979}. But the true value of
$\gamma$ is generally unknown. 

By neglecting the effect of $\Kxc$ in $\partial V/\partial \rho$, 
the Kerker scheme sets $C$ to
\begin{equation}
  C = \alpha \mcF^{-1} D_K \mcF,
  \label{eqn:alphaKerker}
\end{equation}
where $\mcF$ is the matrix representation of the discretized 
Fourier basis that diagonalizes both $v_c$ and $\chi$,
and the diagonal matrix $D_K$ contains $q^2/(q^2+4\pi\hat{\gamma})$ on its 
diagonal, for some appropriately chosen constant $\hat{\gamma}$, and
$\alpha$ is a parameter chosen to ensure~\eqref{eqn:eigbound} is satisfied.

As a result, the eigenvalues of $CJ$ associated with the Kerker
preconditioner are approximately $\alpha(q^2+4\pi \gamma)/(q^2 + 4 \pi \hat{\gamma})$.  
When $q$ is small, the corresponding eigenvalue of $CJ$ is approximately 
$\alpha\gamma/\hat{\gamma}$. When $q$ is large, 
the corresponding eigenvalue of $CJ$ is approximately $\alpha$. 
By choosing an appropriate $\alpha \in (0,1)$ we can ensure that 
all eigenvalues are within $(0,2)$ even when 
$\hat{\gamma}$ is not completely in agreement with the true $\gamma$.

The behavior of $\chi$ for simple insulating systems is very different
from that for simple metallic systems. For simple insulating systems, 
the eigenvalues of $\chi$ corresponding to small $q$ modes behave like
$-\xi q^2$ where $\xi>0$ is a constant~\cite{GhosezGonzeGodby1997,PickCohenMartin1970}. As a result, the spectral radius
of $J$ is bounded by a constant when the contribution from the
exchange-correlation is negligible. Therefore, we can choose $C = \alpha
I$ with an appropriate $\alpha$ to ensure the condition
(\ref{eqn:eigbound}) is satisfied. The optimal choice of $\alpha$ will
be discussed in the next section.

We should also note that when $C_k$ is allowed to change from
one iteration to another through the use of quasi-Newton updates, the 
convergence of the preconditioned fixed point (or quasi-Newton) iteration
can be Q-superlinear~\cite{NocedalWright1999}.

\section{Convergence rate and size dependency}\label{sec:convergence}
In the previous section, we identified the condition under which
a preconditioned fixed-point iteration applied to the Kohn-Sham
problem converges. In this section, we discuss the optimal rate
of convergence and its dependency on the size of the physical system.
Ideally, we would like to construct a preconditioner to ensure
the rate of convergence to be independent of the system size.

\subsection{The convergence rate of the simple mixing scheme}
When the preconditioner is chosen to be $C=\alpha I$ (i.e., simple mixing), 
the convergence of the preconditioned fixed point iteration is 
guaranteed if $\alpha$ satisfies the condition given in (\ref{eqn:alfabond1}).
As is the case for analyzing Richardson's iteration for linear
equations, it is easy
to show that the optimal choice of $\alpha$, which is the solution to
the following problem
\[
r = \min_{\alpha} \max_{\lambda(J_{\ast})} | 1 - \alpha \lambda(J_{\ast}) |,
\]
must satisfy 
\begin{equation}
  \abs{1-\alpha\lambda_{\max}} = \abs{1-\alpha\lambda_{\min}},
  \label{minmaxeig}
\end{equation}
where $\lambda_{\max}$ and $\lambda_{\min}$ are the largest and smallest 
eigenvalues of $J_{\ast}$ respectively.

The solution to (\ref{minmaxeig}) is
\begin{equation}
  \alpha=\frac{2}{\lambda_{\max}+\lambda_{\min}}.
  \label{eqn:midlam}
\end{equation}
Therefore, the optimal convergence rate of simple mixing is
simply~\cite{DederichsZeller1983}
\begin{equation}
  r = \frac{\lambda_{\max}-\lambda_{\min}}{\lambda_{\max}+\lambda_{\min}}
  = \frac{\kappa(J_{\ast})-1}{\kappa(J_{\ast})+1},
\label{eqn:alfarate}
\end{equation}
where $\kappa(J_{\ast})=\lmax / \lmin$ is the condition number of the Jacobian at the solution.

\subsection{The convergence rate of the Anderson/Pulay scheme}

The convergence rate of Broyden's method 
can be shown to be Q-superlinear when it is applied to a smooth 
function, and when the starting guess of the solution is sufficiently close 
to the true solution and the starting guess of Jacobian is sufficiently 
close to the true Jacobian at the solution \cite{NocedalWright1999}. 
However, in the Anderson or Pulay scheme, we reset the previous
approximation to the Jacobian to $C_0 = \alpha I$ at each iteration. 
Therefore, its convergence may not be superlinear in general.  

One interesting observation made by a number of researchers 
\cite{AkaiDederichs1985,DederichsZeller1983} is that $V_{k+1} - \Vstar$ 
approximately lies in
the Krylov subspace 
$\{V_0 - \Vstar, \Jstar (V_0 - \Vstar), ..., \Jstar^{k}(V_0 - \Vstar)\}$
when $V_0$ is sufficiently close to $\Vstar$,
and $V_{k+1}$ is constructed to have a minimum $\|V_{k+1}-\Vstar\|$ in this 
subspace in the Anderson/Pulay scheme even though we do not know this 
subspace explicitly.  (Since we do not know $\Vstar$ or $\Jstar$.)
Therefore, one can draw a connection between the Anderson/Pulay
scheme and the GMRES~\cite{SaadSchultz1986} algorithm for solving 
a linear system of
equations~\cite{FangSaad2009,RohwedderSchneider2011,WalkerNi2009}.  As a
result, if the Anderson or Pulay scheme converges, and when the
pseudo-inverse of $Y_{k}$ is computed in exact arithmetic,
heuristic reasoning suggests that the convergence rate may be
approximately bounded by
\[
  r = \frac{\sqrt{\kappa(J_{\ast})}-1}{\sqrt{\kappa(J_{\ast})}+1}.
\]
Clearly, when $\kappa(J_{\ast})$ is large, the Anderson/Pulay 
acceleration scheme is superior to the simple mixing scheme.

When a good initial approximation to the inverse of the Jacobian (e.g. 
the Kerker preconditioner), $C_0$ is available, it can be combined with 
the Anderson/Pulay acceleration scheme to make the fixed point iteration
converge more rapidly.

%
%

\subsection{The dependency of the convergence rate on system size}
A natural question that arises when we apply a preconditioned fixed
point iteration to a large atomistic system is whether the convergence
rate depends on the size of the system.

For periodic systems, the size of the system is often characterized
by the number of unit cells in the computational domain. To simplify our discussion, we assume
the unit cell to be a simple cubic cell with a lattice constant $L$.
For non-periodic systems such as molecules, we can construct a fictitious
(cubic) supercell that encloses the molecule and periodically extend the 
supercell so that properties of the system can be analyzed through 
Fourier analysis.  In both cases, we assume the number of atoms in 
each supercell is proportional to $L^3$.

Because the convergence rates of both the simple mixing and Anderson's method
depend on the condition number of $\Jstar$, we should examine
the dependency of $\kappa(\Jstar)$ with respect to $L$.  When 
a good initial guess to the Jacobian $C_0$ is available, we should
examine the dependency of $\kappa(C_0\Jstar)$ with respect to $L$.  
Recall that $\Jstar = I - (v_c + \Kxc)\chi$, where $v_c$ is positive 
definite, $\Kxc$ is symmetric but not necessarily positive definite 
and $\chi$ is symmetric negative semidefinite.
The eigenvalues of $\Jstar$ satisfy
$\lambda(\Jstar) > \bar{\lambda} >0$ where $\bar{\lambda}$ is independent 
of the system size.  The inequality $\lmin >\bar{\lambda} > 0$ gives the 
stability condition of the system. 

The dependency of $\lmax$ on $L$ is generally difficult
to analyze. However, for simple model systems such as a jellium system
(or uniform electron gas) in which $\Kxc(\vr,\vr') = \Kxc^{\ast} \delta(\vr,\vr')$ for
some constant $\Kxc^{\ast}$,
we may use Fourier analysis to show that the eigenvalues of $\Jstar$ are 
simply
\begin{equation}
  \lambda_{\vq} = 1 + \left( \frac{4\pi}{q^2} +
  \Kxc^{\ast} \right)  \gamma F_L(q)
  \label{eqn:eigjellium}
\end{equation}
where $q = |\vq|$, $\gamma$ is a constant, and $F_L(q)$ is known as the 
{\em Lindhard} response function~\cite{Ziman1979}. 
The Lindhard function satisfies 
\begin{equation}
  \lim_{q\to 0} F_{L}(q) = 1, \quad \lim_{q\to \infty} F_{L}(q) = 0,
  \label{eqn:lindhardlimit}
\end{equation}
Hence by taking $q=\frac{2\pi}{L}$, $\lmax(\Jstar)$ is determined by
$1+\gamma(L^2/\pi + \Kxc^{\ast})$.  As a result, the convergence of a
fixed point iteration preconditioned by simple mixing and/or modified by
Anderson's method tends to become slower for a jellium system as the
system size increases.  

When the Kerker preconditioner is used, 
the eigenvalues of $C\Jstar$ are 
\begin{equation}
 \lambda_{\vq} = \alpha \frac{q^2+\gamma F_{L}(q) (4\pi +\Kxc^{\ast}q^2)}{q^2 + 4\pi
\hat{\gamma}}.
  \label{}
\end{equation}
They are approximately $\alpha$ when $q$ is large, and are determined by 
$\alpha F_L(q) \gamma/\hat{\gamma}$ when $q$ is small.  Since the smallest 
$q$ satisfies $q = 2\pi/L$, 
the convergence rate of the Kerker preconditioned fixed point iteration is
independent of system size for a jellium system.  The same conclusion
can be reached for simple metals such as Na or Al which behave like free
electrons~\cite{Ziman1979}.  Therefore, the Kerker preconditioner is an
ideal preconditioner for simple metals.

However, the Kerker preconditioner is not an appropriate preconditioner 
for insulating systems.
Although in general the Jacobian associated
with the insulating system cannot be diagonalized by the Fourier basis,
it can be shown that $e^{i \vq \cdot \vr}$ is an approximate eigenfunction of 
$\chi$ with the corresponding eigenvalue $-\xi
q^2$~\cite{GhosezGonzeGodby1997,PickCohenMartin1970}. If we neglect
the contribution 
from $\Kxc$, $e^{i \vq \cdot \vr}$ is also an approximate eigenfunction 
of $\Jstar$ with the corresponding eigenvalue 
$1 + (4\pi/q^2) q^2 \xi= 1+4\pi \xi$ for small $q$'s.  
If $C$ is chosen to be the Kerker preconditioner, then the corresponding 
eigenvalue of $C\Jstar$ is 
\[
\lambda_{\vq} = \frac{q^2}{q^2 + 4\pi \hat{\gamma}} (1 + 4 \pi \xi).
\]
As the system size $L$ increases, the smallest $q$, which satisfies
$q = 2\pi/L$, becomes smaller.  Consequently, the corresponding eigenvalue of 
$C\Jstar$ approaches zero. The convergence rate, which is determined 
by $\sigma(1-C\Jstar)$, deteriorates as the system size increases.

For insulating systems, a good preconditioner is simply $\alpha I$, where
$\alpha$ is chosen to be close to $1/(1+ 4\pi \xi)$ (in general, we do 
not know the value of $\xi$). When such a preconditioner is used
the convergence the fixed point iteration becomes independent of 
the system size.

\section{Elliptic preconditioner}\label{sec:elliptic}
As we have seen above, simple insulating and metallic systems call for different
types of preconditioners to accelerate the convergence of a fixed point
iteration for solving the Kohn-Sham problem. A natural question 
one may ask is how we should construct a preconditioner for 
a complex material that may contain both insulating and metallic components
or metal surfaces.

Before we answer this question, we should point out that the analysis
of the spectral properties of $\Jstar$ that we presented earlier 
relies heavily on the assumption that the eigenfunctions of $\Jstar$
are approximately planewaves.  Although this assumption is generally 
acceptable for simple materials, it may not hold for more complex
systems.  Therefore, to develop a more general technique for constructing 
a good preconditioner, it may be more advantageous to explore
ways to approximate $\Jstar^{-1}$ or the solution to the equation
$\Jstar \tilde{r}_{k} = r_{k}$ directly for some residue $r_{k}$.

One of the difficulties with this approach is in getting a
good approximation of the polarizability operator $\chi$ in
$\Jstar = I - (v_c + K_{\xc})\chi$. The use of the Adler-Wiser
formula given in~\eqref{eqn:alderwiser} would require computing almost 
all eigenpairs of $H$.  Even when some of the $\psi_m$'s can be replaced 
by simpler functions such as planewaves~\cite{AngladeGonze2008}, constructing this operator and 
working with it would take at least $\mathcal{O}(N^4)$ operations.

Therefore it is desirable to replace the Adler-Wiser representation of
$\chi$ with something much simpler and cheaper to compute. However, 
making such a modification to $\chi$ only may introduce undesirable
error near low electron density regions because $K_{\xc}$ contains terms 
proportional to $\rho^{-2/3}$ (which originates from the Dirac exchange
term \cite{dirac}).  This problem can be avoided by 
using the observations made in the physics community that the 
product of $K_{\xc}$ and $\chi$ is relatively small compared to 
$v_c \chi$, even though this observation has not been rigorously proved.
As a result, it is reasonable to approximate $\Jstar$ by
$\tJstar = I - v_c \chi$. For historical reasons, this approximation is 
known as the random phase approximation (RPA) in the physics 
literature~\cite{AngladeGonze2008}.

Note that, under RPA, we may rewrite $\tJstar^{-1}$ as
\[
\tJstar^{-1}  = (v_c^{-1} - \chi)^{-1} v_c^{-1}.
\]
Since $v_c^{-1} = -\Delta /(4\pi)$, applying
$\tJstar^{-1}$ to a vector $r_k$ 
simply amounts to solving the following equation
\begin{equation}
  (-\Delta-4\pi\chi)\tilde{r}_k = -\Delta r_k.
  \label{eqn:RPAprecond2}
\end{equation}

To construct a preconditioner $C$, we will replace $\chi$ with a
simpler operator. In many cases, we can choose the approximation
to be a local (diagonal) operator defined by a function $b(\vr)$,
although other type of more sophisticated operators are possible.
To compensate for
the simplification
of $\chi$, we replace the Laplacian operator on the left 
of~\eqref{eqn:RPAprecond2} 
by $-\nabla\cdot \left( a(\vr)\nabla \right)$ for some appropriately
chosen function $a(\vr)$. This additional change yields the following 
elliptic partial differential equation (PDE)
\begin{equation}
  \left( -\nabla\cdot \left( a(\vr)\nabla \right) 
  + 4\pi b(\vr)\right) \tilde{r}_{k} = -\Delta r_k.
  \label{eqn:RPAelliptic}
\end{equation}
Because our construction of the preconditioner involves solving
an elliptic equation, we call such a preconditioner an {\em elliptic
preconditioner}.

Although the new framework we use to construct a preconditioner for
the fixed point iteration is based on heuristics and certain simplifications
of the Jacobian, it is consistent with the existing preconditioners that
are known to work well with simple metals or insulators.

For example, for metallic systems, setting $a(\vr)=1$ and 
$b(\vr) =  -\hat{\gamma}$ for some constant $\hat{\gamma}>0$ yields
\begin{equation}
  (-\Delta+4\pi \hat{\gamma})\tilde{r}_k = -\Delta r_k.
  \label{eqn:RPAkerker}
\end{equation}
The solution of the above equation is exactly the same as what is produced by the 
Kerker preconditioner.

For isotropic insulating system, setting $a(\vr)=1+4\pi \xi$ and $b(\vr)=0$
yields 
\[
  -(1+4\pi\xi)\Delta\tilde{r}_k = -\Delta r_k.
\]
The solution to the above equation is simply
\begin{equation}
  \tilde{r}_k = \frac{1}{1+4\pi\xi} r_k.
  \label{eqn:RPAinsulator}
\end{equation}
Such a solution corresponds to simple mixing with $\alpha$ set to
$1/(1+4\pi \xi)$.

For a complex material that consists of both insulating and metallic
components, it is desirable to choose approximation of $a(\vr)$ and 
$b(\vr)$ that are spatially dependent. 
The asymptotic behavior of $\chi$ with respect to the sizes of both insulating 
and metallic systems suggests that  $a(\vr)$ and $b(\vr)$ should be chosen to 
satisfy $a(\vr)\ge 1$ and $b(\vr)\ge 0$.  In this case, the operator defined on the 
left hand side of (\ref{eqn:RPAelliptic}) is a strongly elliptic operator. Such 
an operator is symmetric positive semi-definite.  

The implementation of the elliptic preconditioner only requires solving
an elliptic equation.
In general $a(\vr), b(\vr)$ are spatially dependent, and solving the
elliptic preconditioner requires more than just a Fourier transform
and scaling operation as is the case for the Kerker preconditioner.  
However, it is generally much less time consuming than 
evaluating the Kohn-Sham map or constructing $\chi$
or $J_k$.  In particular, fast algorithms such as
multigrid~\cite{Brandt1977}, 
fast multipole method (FMM)~\cite{GreengardRokhlin1987}, Hierarchical
matrix~\cite{Hackbusch1999} solver 
and Hierarchical semi-separable (HSS)
matrix~\cite{ChandrasekaranGuPals2006} can be 
applied to solve Eq.~\eqref{eqn:RPAelliptic} with $\Or(N)$ arithmetic 
operations.  Even if we cannot achieve $\Or(N)$ complexity, 
Eq.~\eqref{eqn:RPAelliptic} can often be solved efficiently by 
Krylov subspaces iterative methods as we will show in the next section.

Our numerical experience suggests that simple choices of $a(\vr)$ and $b(\vr)$ 
can produce satisfactory convergence result for complicated systems.  For
example, if we place a metallic system in vacuum to ascertain its 
surface properties~\cite{ShollSteckel2009}, we can choose $b(\vr)$ to be a nonzero 
constant in the metallic region, and almost $0$ in the vacuum part.  
The resulting piecewise constant function can be smoothed by convolving 
it with a Gaussian kernel.  Similarly, $a(\vr)$ can be chosen to be $1$ 
in the metallic region, and a constant larger than $1$ in the vacuum
region.

Due to the simplification that we made about the $\chi$ term in the Jacobian
and the omission of the $K_{\xc} \chi$ term altogether, the construction of 
an elliptic preconditioner alone may not be sufficient to reduce 
the number of fixed point iterations required to reach convergence.
However, such a preconditioner can be easily combined with 
the Broyden type of quasi-Newton technique such as Anderson's method discussed
in~\ref{sec:broyden} to further improve the convergence of the 
self-consistent field iteration. This is the approach we take in
the examples that we will show in the next section.

\section{Numerical results}\label{sec:numerical}
In this section, we demonstrate the performance of the elliptic 
preconditioner proposed in the previous section, and compare it with
other acceleration schemes through two examples. 
The first example consists of a one-dimensional (1D) reduced
Hartree-Fock model problem that 
can be tuned to exhibit both metallic and insulating features. 
The second example is a three-dimensional (3D)
problem we construct and solve in KSSOLV~\cite{YangMezaLeeEtAl2009},
which is a MATLAB toolbox for solving Kohn-Sham equations for small
molecules and solids implemented entirely in MATLAB m-files. KSSOLV 
uses planewave expansion to discretize the Kohn-Sham equations. 
It also uses the Troullier-Martins pseudopotential~\cite{TroullierMartins1991}
with the LDA exchange-correlation functional to approximate the ionic
potential.

\subsection{One dimensional reduced Hartree-Fock model}

The 1D reduced Hartree-Fock model was introduced by Solovej~\cite{Solovej1991}, 
and has been used for analyzing defects in solids
in~\cite{CancesDeleurenceLewin2008,CancesDeleurenceLewin2008a}. The simplified
1D model neglects the contribution of the exchange-correlation term.  
Nonetheless, typical behaviors of an SCF iteration observed for 3D problems 
can be exemplified by this 1D model. In addition to neglecting the 
exchange-correlation potential, we also use a pseudopotential to 
represent the electron-ion interaction. This makes our 1D model 
slightly different from that presented in \cite{Solovej1991}.

The Hamiltonian in our 1D reduced Hartree-Fock model is given by
\begin{equation}
   H[\rho] = -\frac{1}{2} \frac{d^2}{dx^2} + \int K(x,y) (\rho(y)+m(y)) \ud y
   \label{eqn:HrHF}
\end{equation}
Here $m(x)=\sum_{i=1}^{M} m_i(x-R_i)$, with the position of the
$i$-th nuclei denoted by $R_{i}$.  Each function $m_{i}(x)$ takes the
form
\begin{equation} 
  m_{i}(x) = -\frac{Z_i}{\sqrt{2\pi\sigma_{i}^2}}
  e^{-\frac{x^2}{2\sigma_i^2}},
  \label{}
\end{equation}
where $Z_i$ is an integer representing the charge of the $i$-th nucleus.
The parameter $\sigma_{i}$ represents the width of the nuclei in the 
pseudopotential theory.  Clearly as $\sigma_{i}\to 0$, 
$m_{i}(x)\to -Z_{i}\delta(x)$ which is the
charge density for an ideal nucleus. In our numerical simulation, we set
$\sigma_{i}$ to a finite value. The corresponding $m_{i}(x)$ is
called a \textit{pseudo charge density} for the $i$-th nucleus. We
refer to the function $m(x)$ as the total pseudo-charge density of the nuclei.  
The system satisfies charge
neutrality condition, \ie
\begin{equation}
  \int \rho(x) + m(x) \ud x = 0.
  \label{eqn:chargeneutral}
\end{equation}
Since $\int m_i(x) \ud x = -Z_i$, the charge neutrality
condition~\eqref{eqn:chargeneutral} implies 
\begin{equation}
  \int \rho(x) \ud x = \sum_{i=1}^{M} Z_{i} = N,  
  \label{}
\end{equation}
where $N$ is the total number of electrons in the system. To
simplify discussion, we omit the spin contribution here.

Instead of using a bare Coulomb interaction, which diverges in 
1D, we adopt a Yukawa kernel 
\begin{equation}
  K(x,y) = \frac{2\pi e^{-\kappa \abs{x-y}}}{\kappa\epsilon_{0}},
  \label{eqn:YukawaK1}
\end{equation}
which satisfies the equation
\begin{equation}
  -\frac{d^2}{d x^2} K(x,y) + \kappa^2 K(x,y) = \frac{4\pi}{\epsilon_0} \delta(x-y).
  \label{eqn:YukawaK2}
\end{equation}
As $\kappa\to 0$, the Yukawa kernel approaches the bare Coulomb interaction
given by the Poisson equation. The parameter $\epsilon_0$ is used to 
make the magnitude of the electron static contribution comparable to
that of the
kinetic energy.  

The parameters used in the reduced Hartree-Fock model are chosen as
follows.  Atomic units are used throughout the discussion unless
otherwise mentioned.  For all the systems tested below, the distance
between each atom and its nearest neighbor is set to $10$ a.u..  The
Yukawa parameter $\kappa=0.01$ is small enough so that the range
of the electrostatic interaction is sufficiently long, and $\epsilon_0$
is set to $10.00$.  The nuclear charge $Z_{i}$ is set to $2$ for all atoms.  
Since spin is neglected, $Z_{i}=2$ implies that each atom contributes to $2$
occupied bands.  The Hamiltonian operator is represented in a planewave
basis set. 
The temperature of the
system is set to $100$ K, which is usually considered to be very low,
especially for the simulation of metallic systems.

\begin{figure}[h]
  \begin{center}
    \subfloat[]{\includegraphics[width=0.30\textwidth,height=0.30\textwidth]{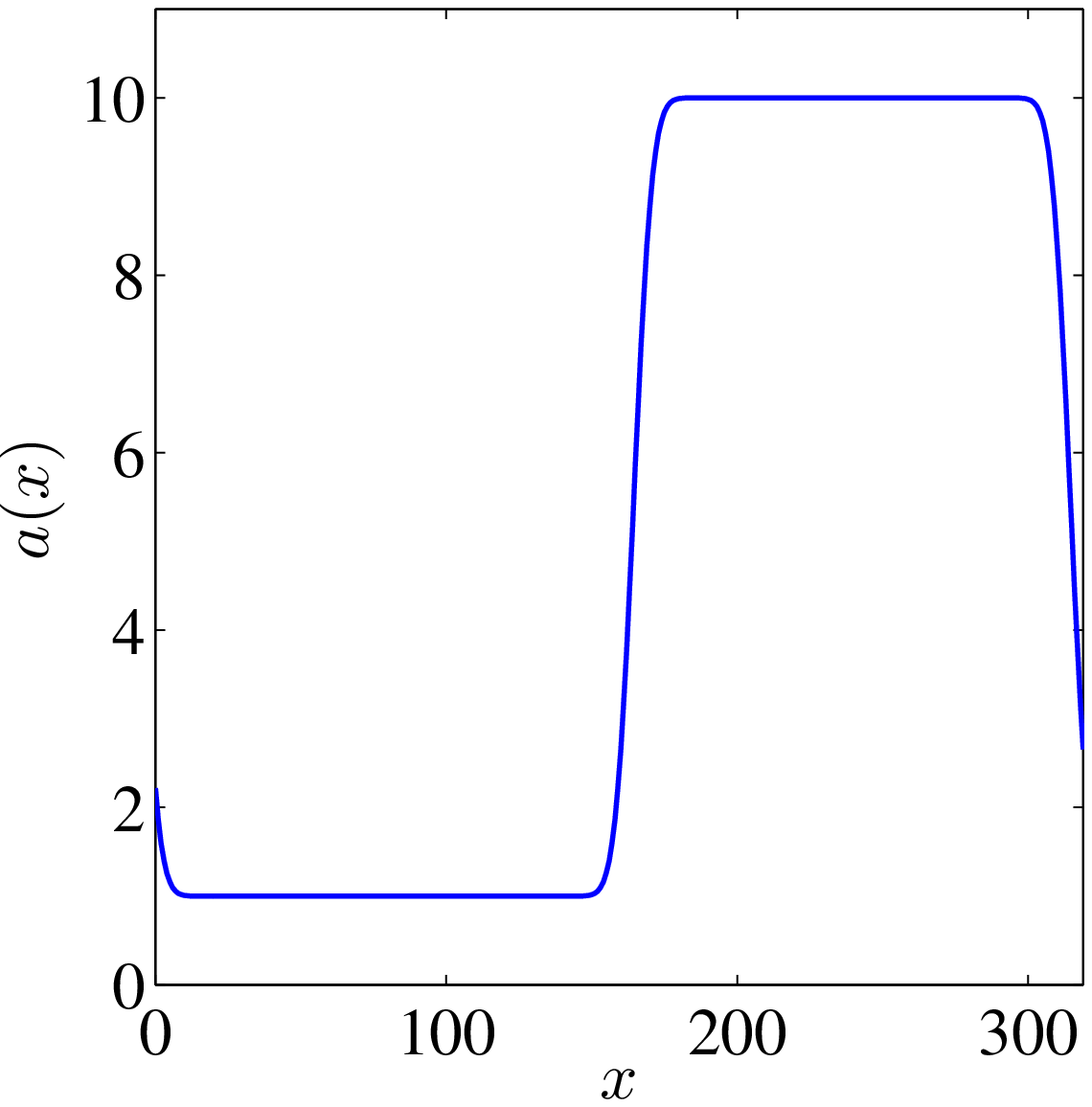}}
    \qquad
    \subfloat[]{\includegraphics[width=0.30\textwidth,height=0.30\textwidth]{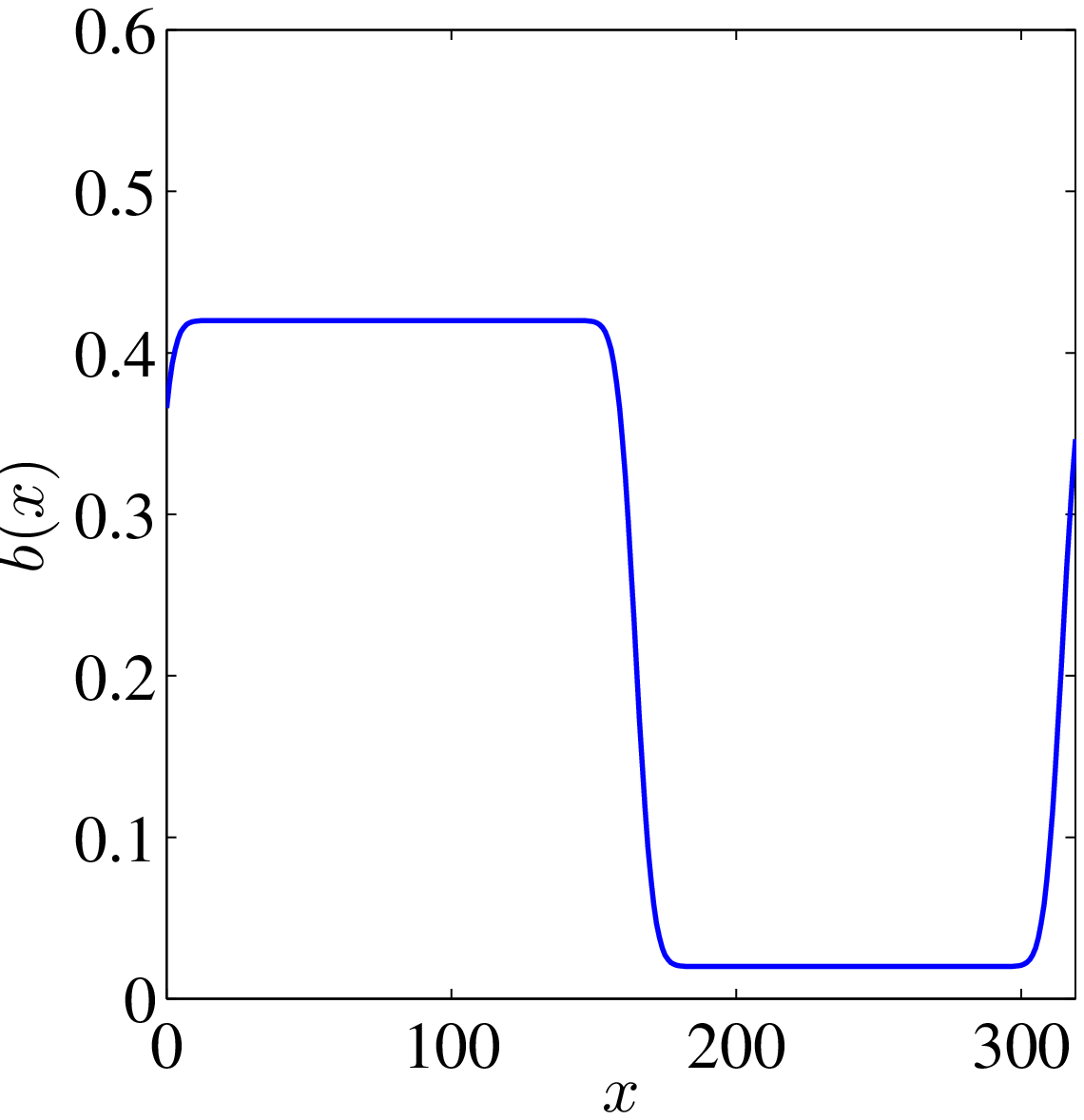}}
  \end{center}
  \caption{(a) The choice of $a(x)$ and (b) the choice of $b(x)$ used by the elliptic
  preconditioner for a system with mixed metallic and insulating region.} 
  \label{fig:axbx1d}
\end{figure}

By adjusting the parameters $\{\sigma_{i}\}$, the reduced Hartree-Fock
model can be tuned to resemble an insulating, metallic or  hybrid 
system.  We apply the elliptic preconditioner with different
choices of $a(x)$ and $b(x)$ to all three cases. In 
the case of an insulator and a metal, both $a(x)$ and $b(x)$ are 
chosen to be constant functions. For the hybrid system, $a(x)$ and
$b(x)$ are constructed by convolving a step function with a Gaussian
kernel as shown in Figure~\ref{fig:axbx1d}.  The $\sigma_i$
values used for all these cases are listed in Table~\ref{tab:prob1}
along with the constant values chosen for $a(x)$ and $b(x)$ in
the insulating and metallic cases.  For the hybrid case, we partition
the entire domain $[0,320]$ into two subdomains: $[0,160]$ and $[160,320]$.
The $\sigma_i$ value is set to $6.0$ in the first subdomain and 
$2.0$ in the second subdomain.
\begin{table}[htbp]
\center
\begin{tabular}{|c|c|c|c|c|} \hline
 case        & $\sigma_i$ & $a(x)$ & $b(x)$ &  $\hat{\gamma}$ \\ \hline
 insulating  &  2.0       &  1.0   &   0.0  &   0.50  \\ \hline
 metallic    &  6.0       &  1.0   &   0.5  &   0.50  \\ \hline
 hybrid      & 2.0/6.0   &  \mbox{see Fig.~\ref{fig:axbx1d} (a)}   &
 \mbox{see Fig.~\ref{fig:axbx1d} (b)} &   0.42 \\
\hline
\end{tabular}
\caption{Test cases and SCF parameters used for the 1D model.}
\label{tab:prob1}
\end{table}

For all three cases, we apply Anderson's method,
Anderson's method combined with the Kerker preconditioner, and
Anderson's method combined with the elliptic preconditioner to
the SCF iteration.  The $\alpha$ parameter used in
the Anderson scheme is set to $0.50$ in all tests.  The $\hat{\gamma}$
parameter is set to $0.50$ for the insulating and metallic cases, and
$0.42$ for the hybrid case.

\begin{figure}[h]
  \begin{center}
    \subfloat[Insulator]{\includegraphics[width=0.30\textwidth,height=0.30\textwidth]{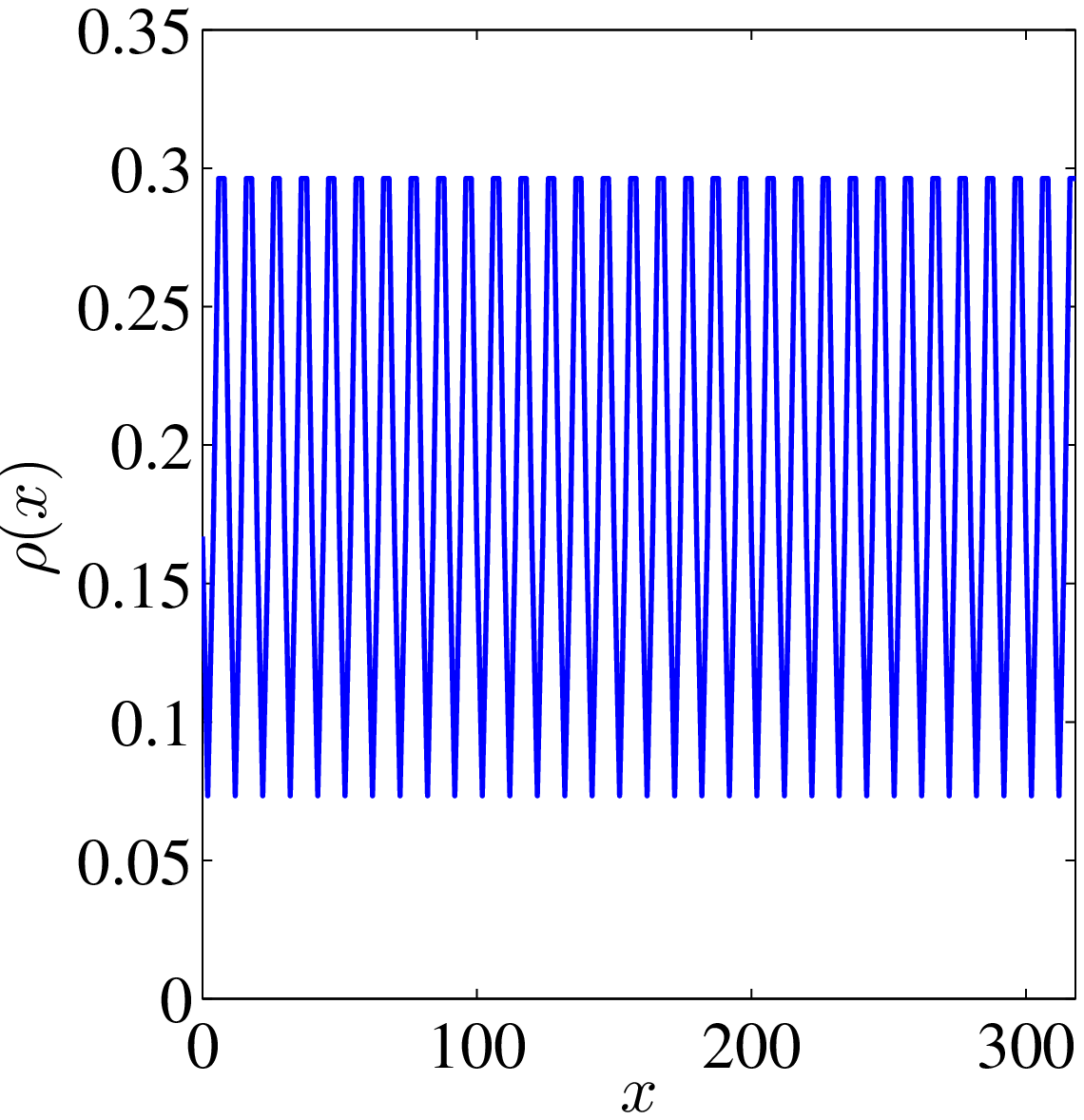}}
    \qquad
    \subfloat[Insulator]{\includegraphics[width=0.30\textwidth,height=0.30\textwidth]{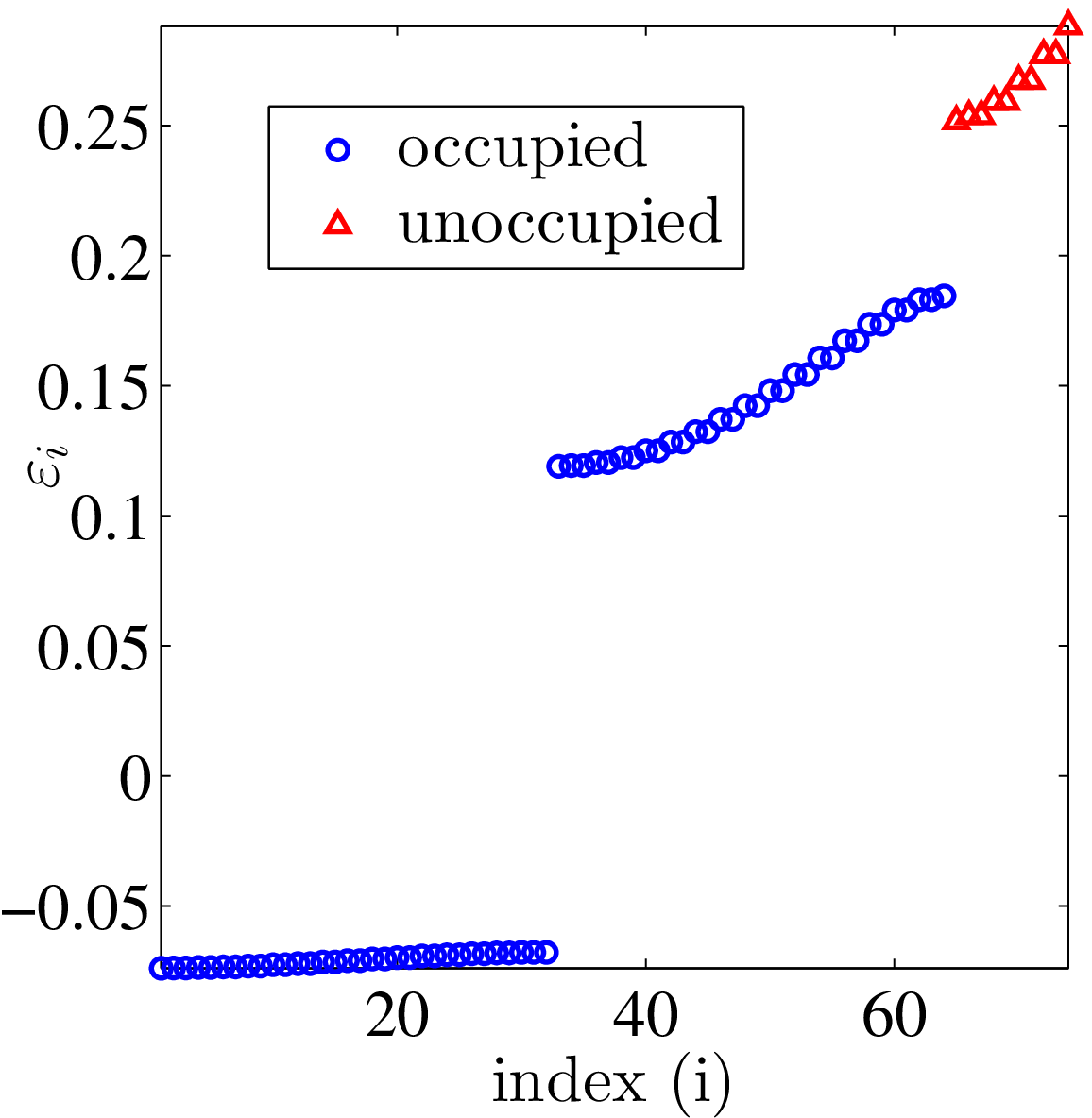}} \\
    \subfloat[Metal]{\includegraphics[width=0.30\textwidth,height=0.30\textwidth]{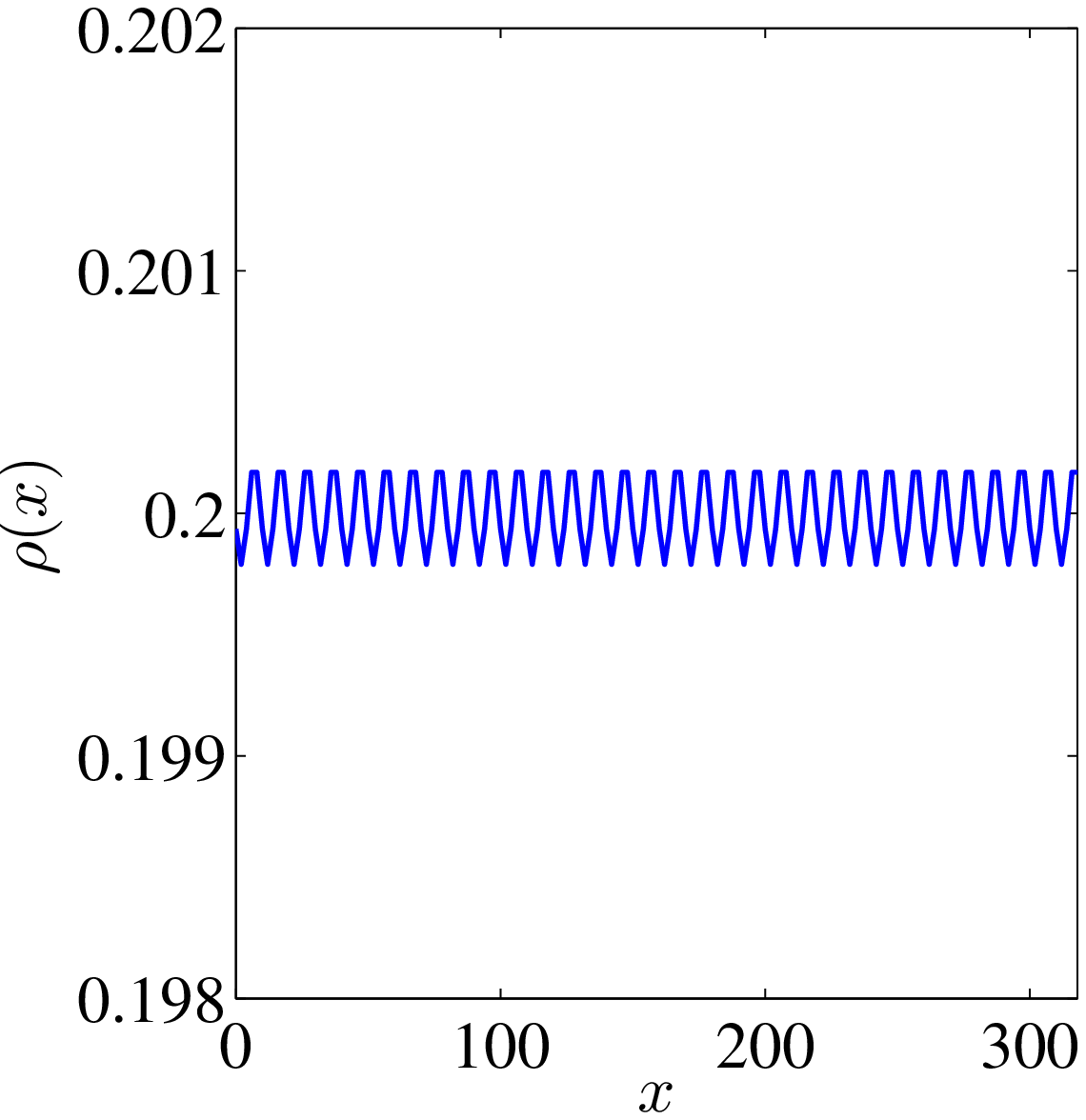}}
    \qquad
    \subfloat[Metal]{\includegraphics[width=0.30\textwidth,height=0.30\textwidth]{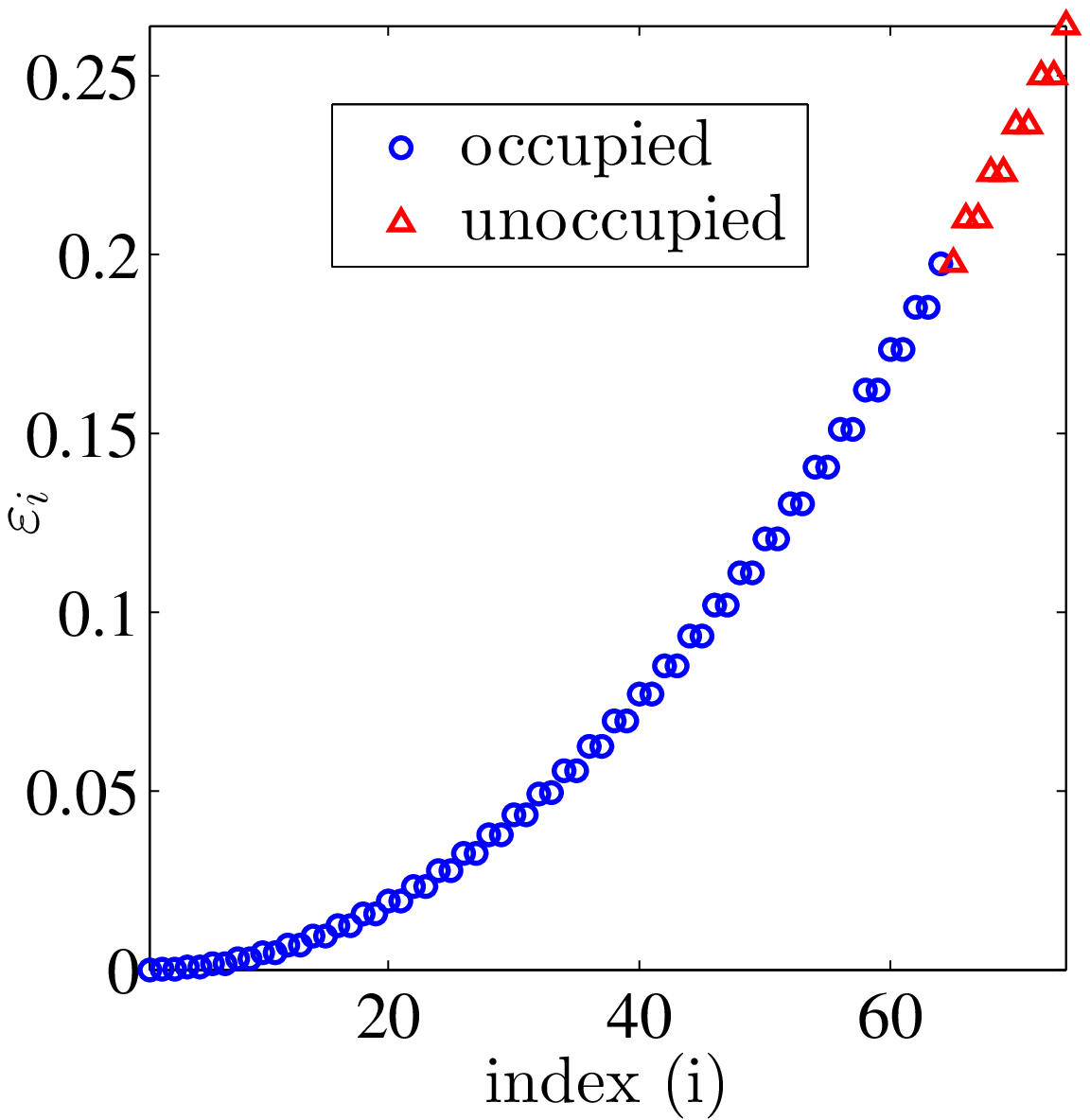}}\\
    \subfloat[Metal+Insulator]{\includegraphics[width=0.30\textwidth,height=0.30\textwidth]{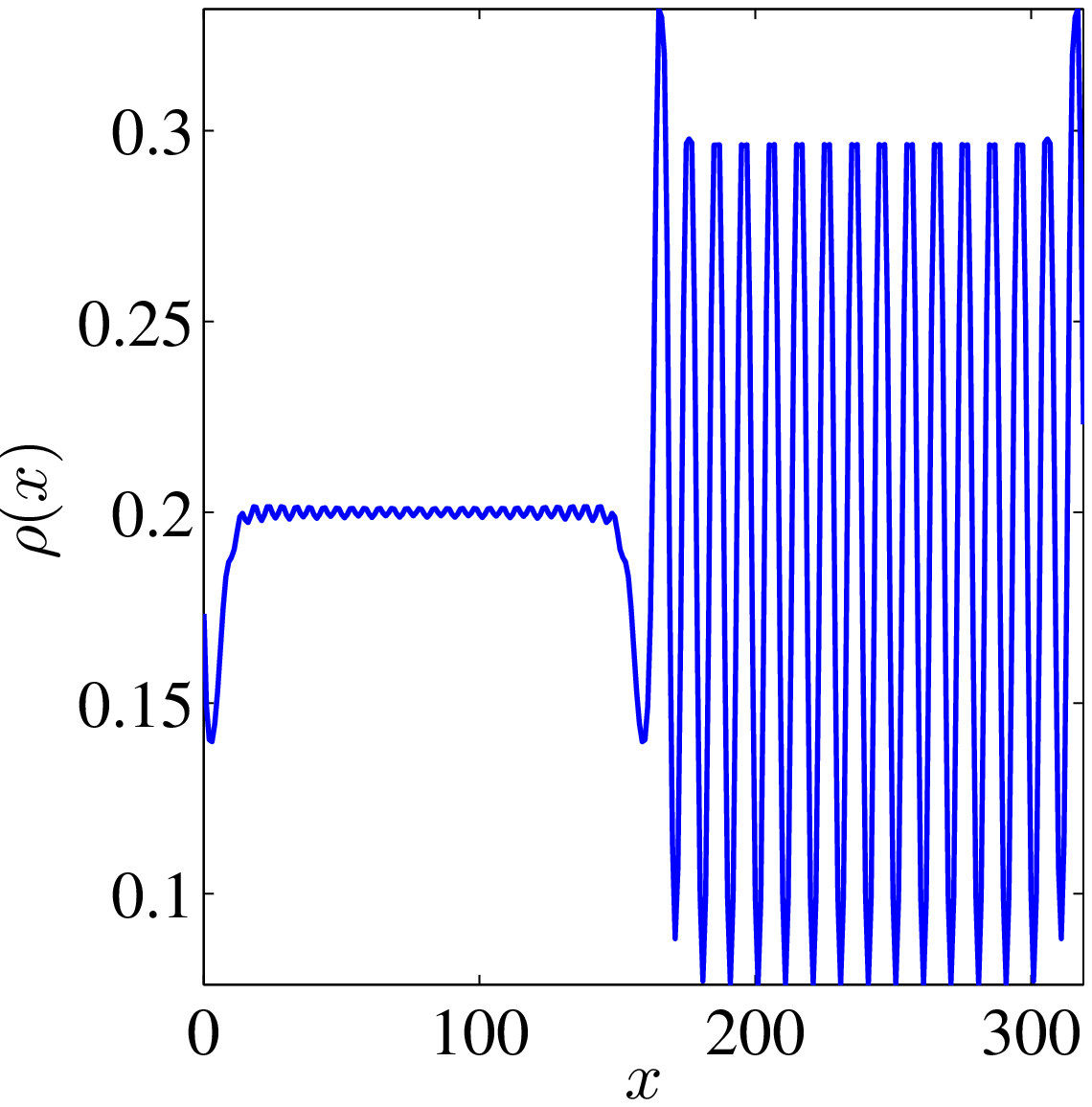}}
    \qquad
    \subfloat[Metal+Insulator]{\includegraphics[width=0.30\textwidth,height=0.30\textwidth]{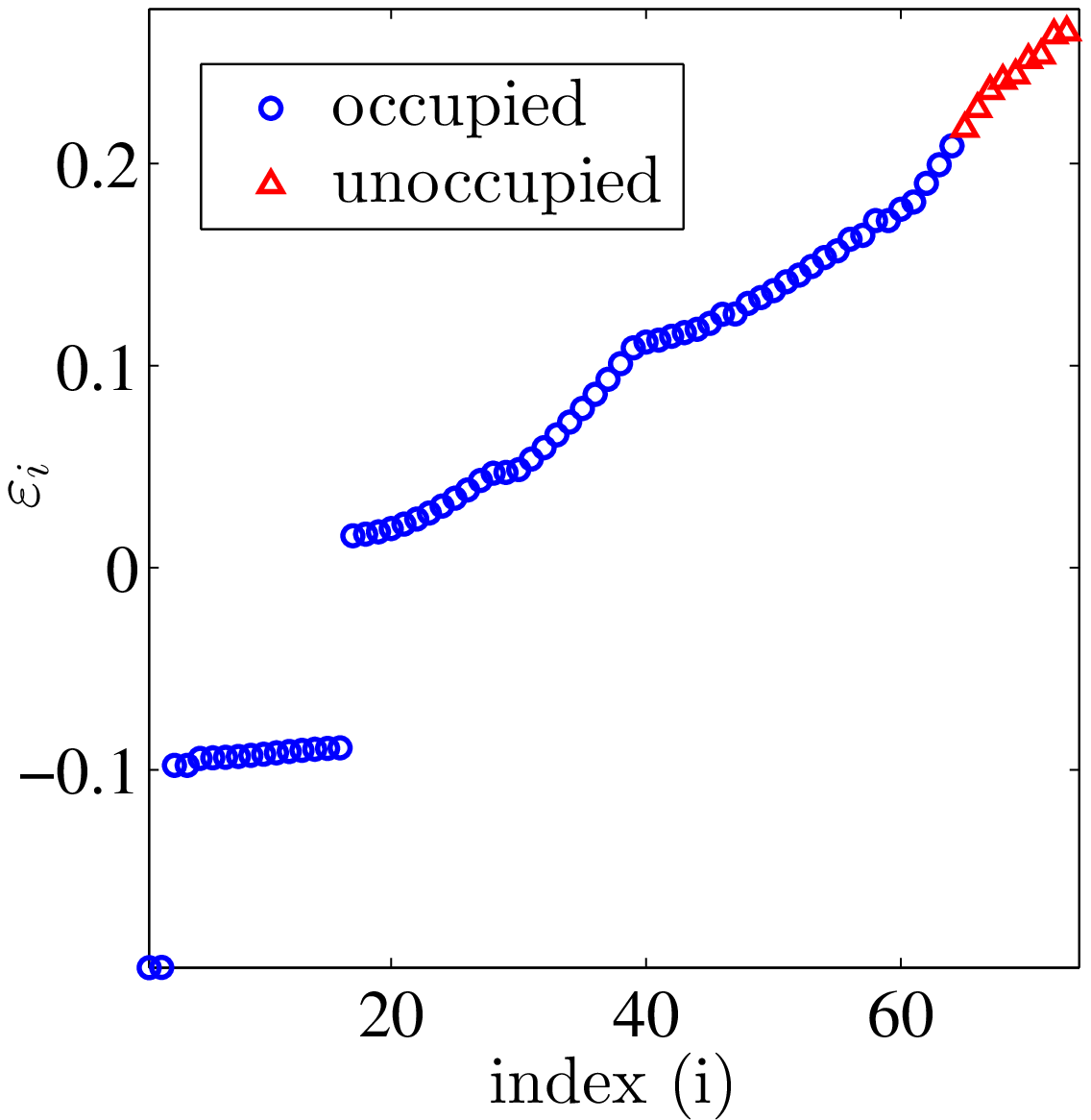}}
\end{center}
  \caption{The electron density $\rho(x)$ of a 32-atom (a) insulating system  
  (c) metallic and (e) hybrid metal-insulator in the left panel. 
  The corresponding occupied (blue circles) and unoccupied eigenvalues
  (red triangles) are shown in the right panel in subfigure (b), (d), (f),
  respectively.} 
  \label{fig:rhoandeig1d}
\end{figure}

The converged electron density $\rho$ associated with the three 1D test cases 
as well as the 74 smallest eigenvalues associated with the Hamiltonian
defined by the converged $\rho$ are shown in Figure~\ref{fig:rhoandeig1d}. 
The first $64$ eigenvalues correspond to occupied states, and the 
rest correspond to the first $10$ unoccupied states.

For the insulator case, the electron density fluctuates between $0.08$ and
$0.30$.  There is a finite gap between the highest occupied
eigenvalue ($\varepsilon_{64}$) and the lowest unoccupied eigenvalue
($\varepsilon_{65}$). The band gap is $E_{g}=\varepsilon_{65}-\varepsilon_{64}=0.067$ a.u..  The electron density associated with the metallic case 
is relatively uniform in the entire domain. The corresponding eigenvalues 
lie on a parabola (which is the correct distribution for uniform electron gas.)
In this case, there is no gap between the occupied eigenvalues and 
the unoccupied eigenvalues.  For the hybrid case, the electron density 
is uniformly close to a constant in the metallic region (except at the 
boundary), and fluctuates in the insulating region. There is no gap between
the occupied and unoccupied states.
%

\begin{figure}[h]
  \begin{center}
    \subfloat[Insulator]{\includegraphics[width=0.3\textwidth,height=0.3\textwidth]{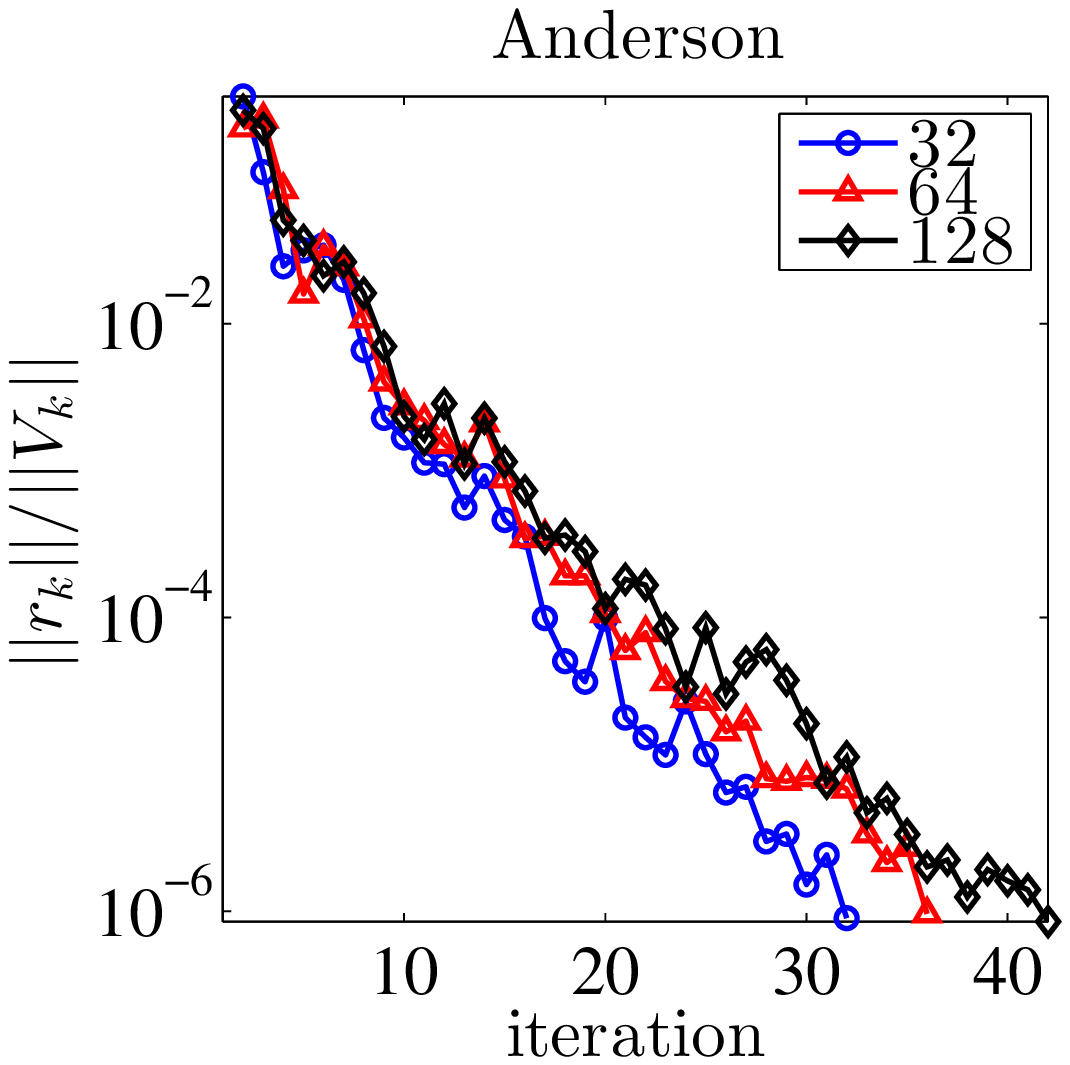}}
    \subfloat[Insulator]{\includegraphics[width=0.3\textwidth,height=0.3\textwidth]{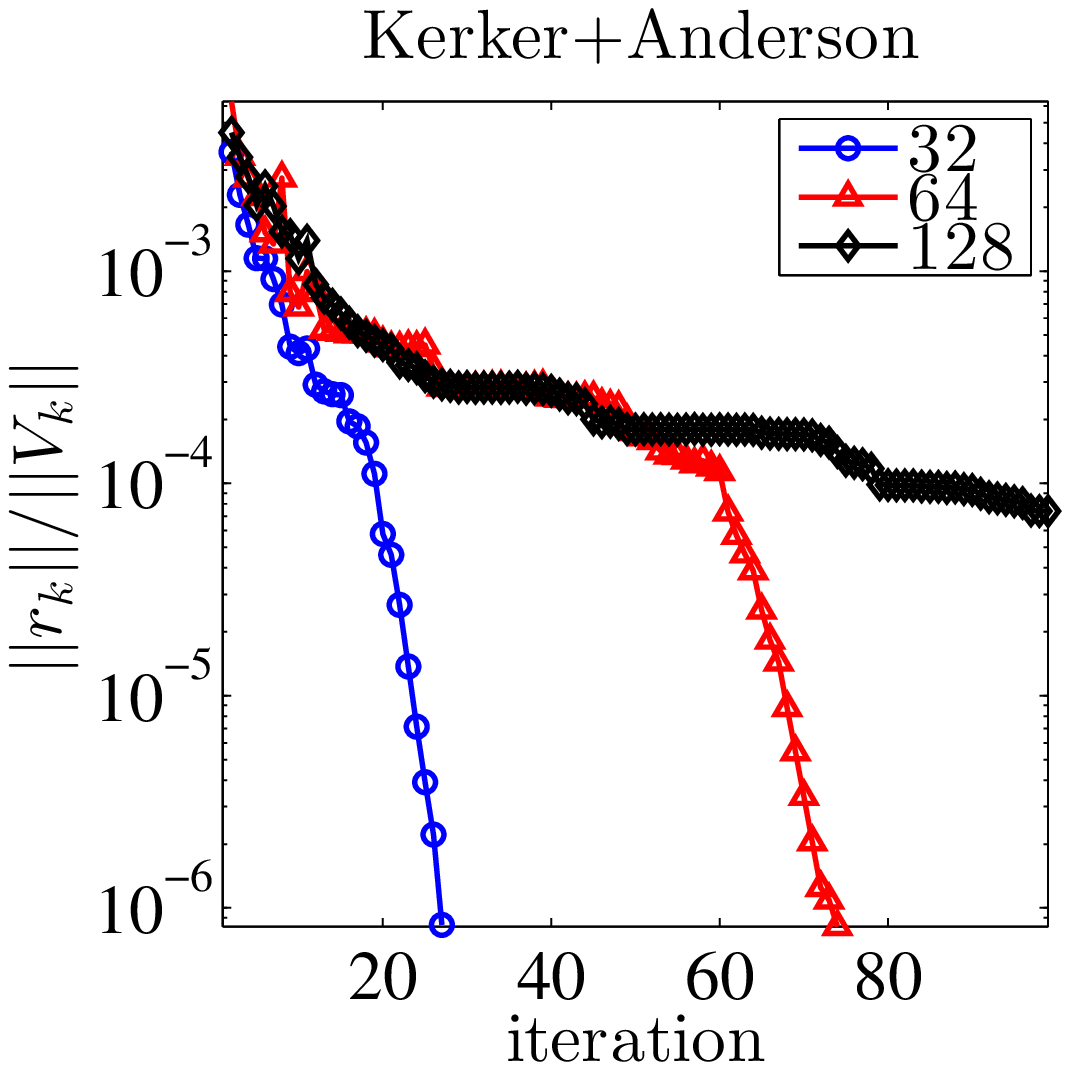}}
    \subfloat[Insulator]{\includegraphics[width=0.3\textwidth,height=0.3\textwidth]{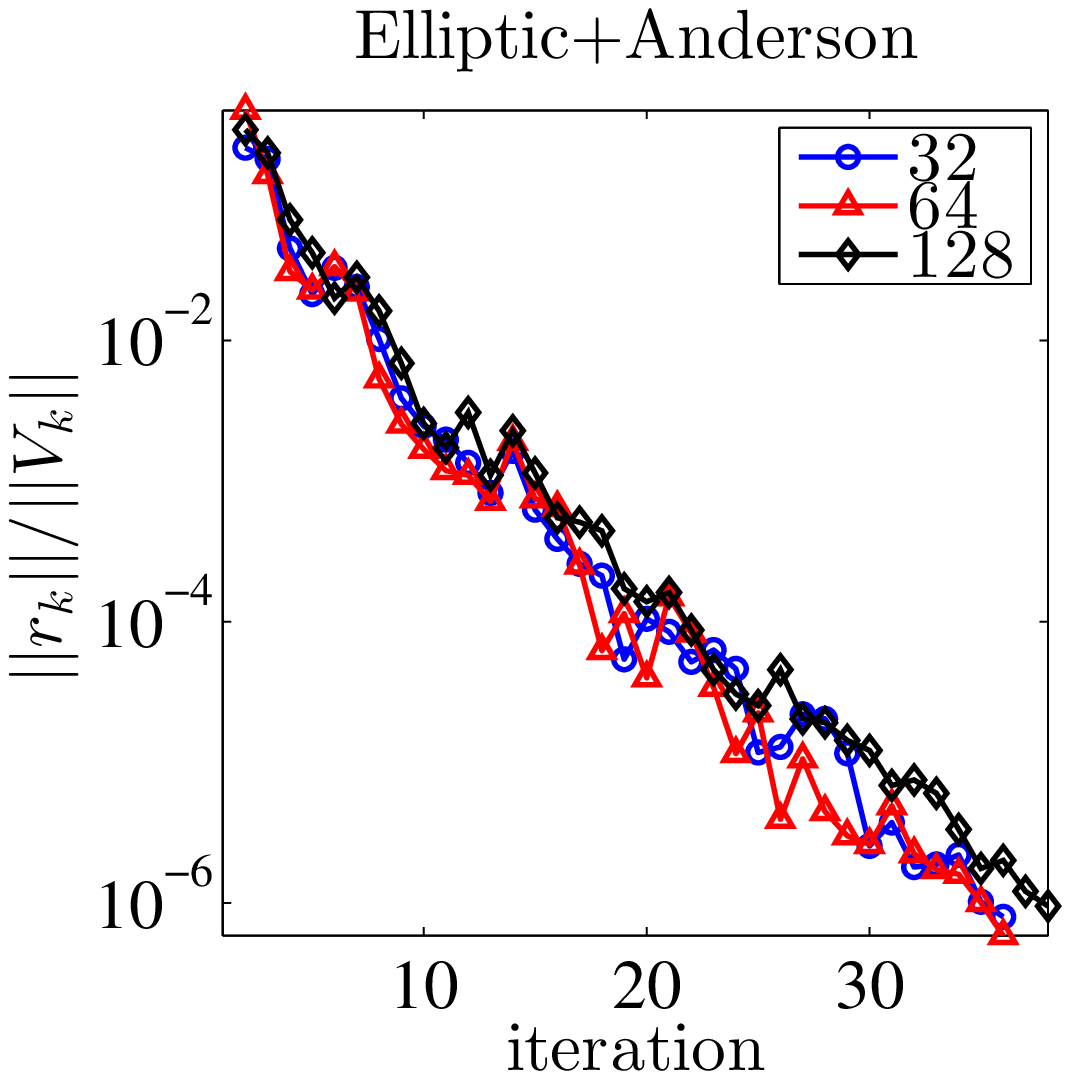}}\\
    \subfloat[Metal]{\includegraphics[width=0.3\textwidth,height=0.3\textwidth]{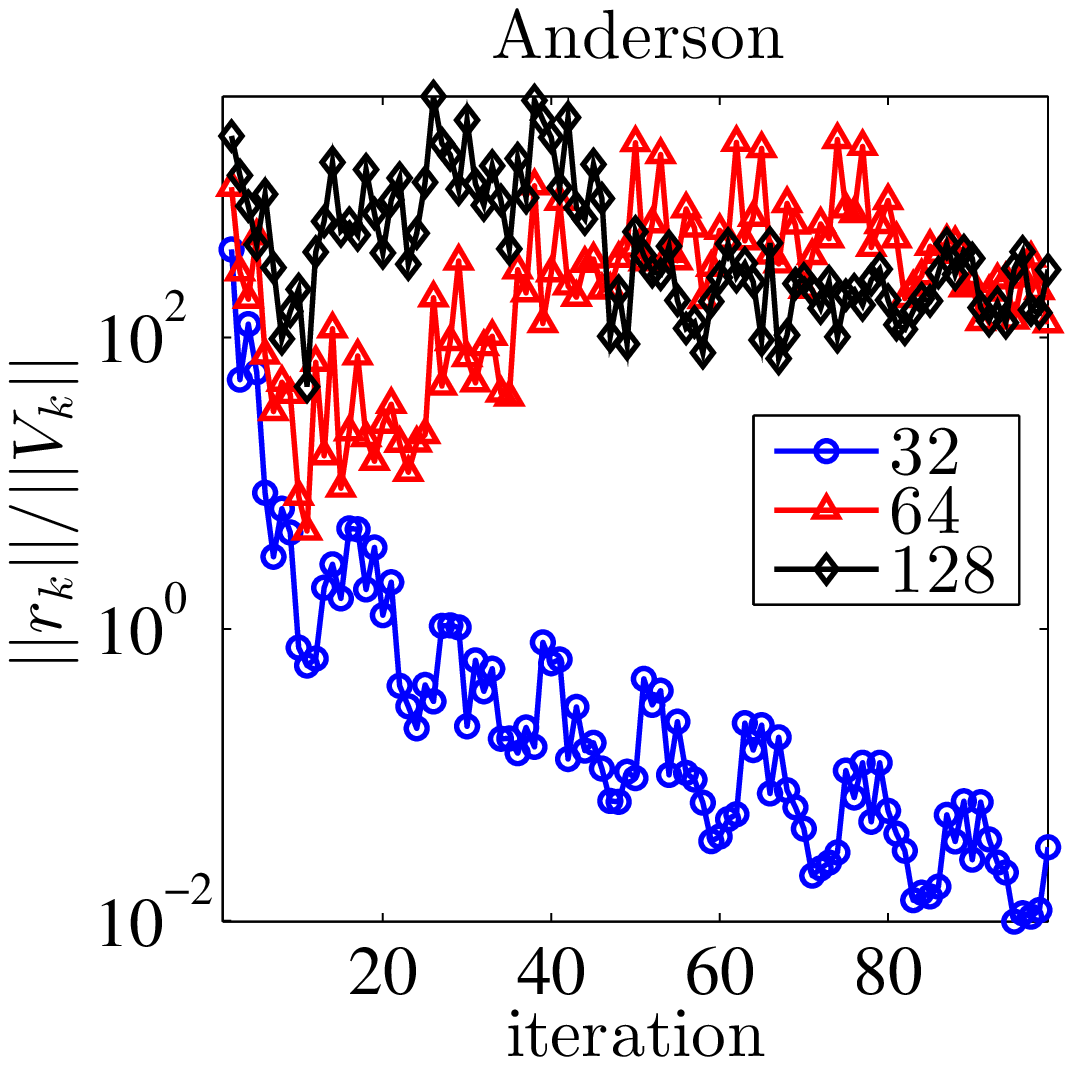}}
    \subfloat[Metal]{\includegraphics[width=0.3\textwidth,height=0.3\textwidth]{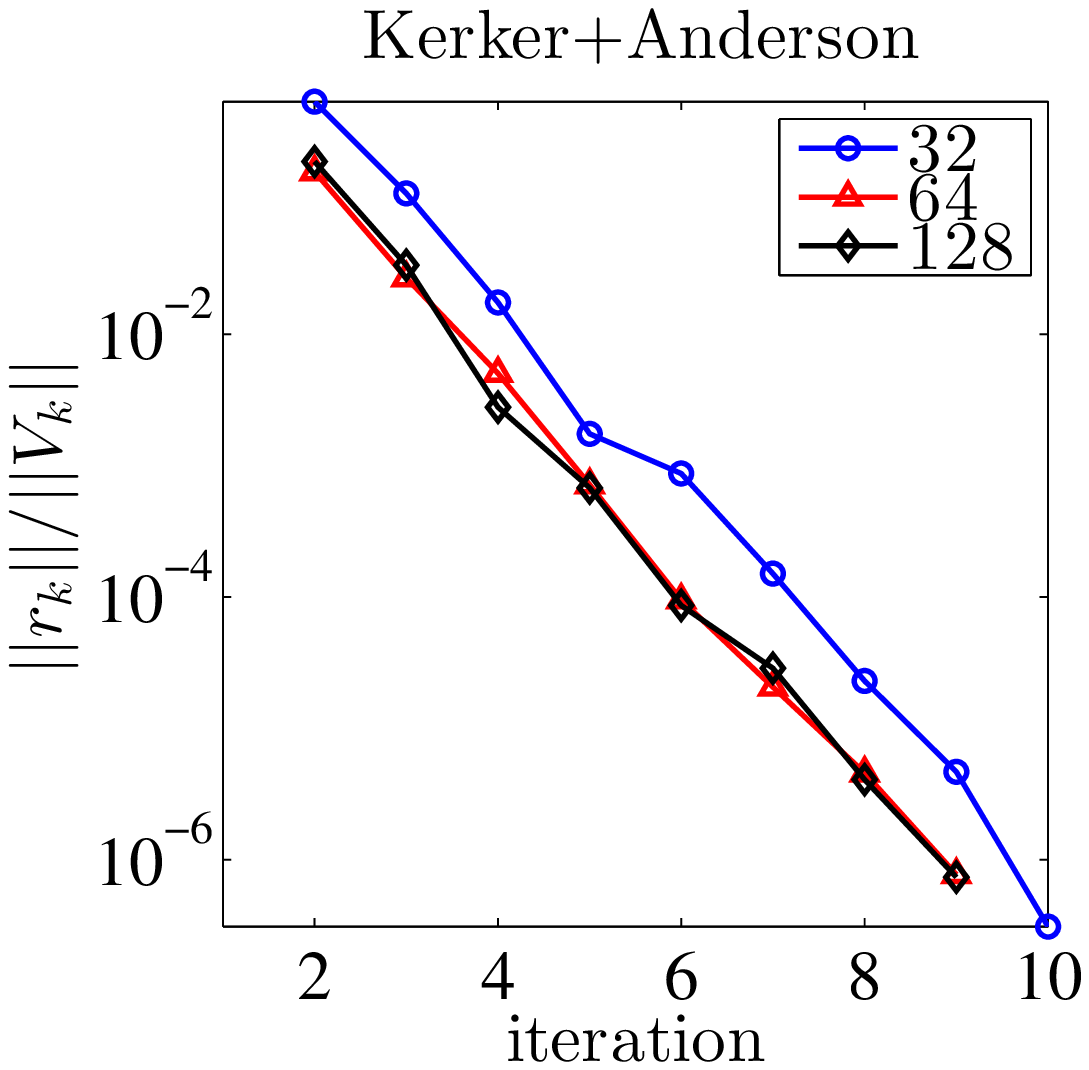}}
    \subfloat[Metal]{\includegraphics[width=0.3\textwidth,height=0.3\textwidth]{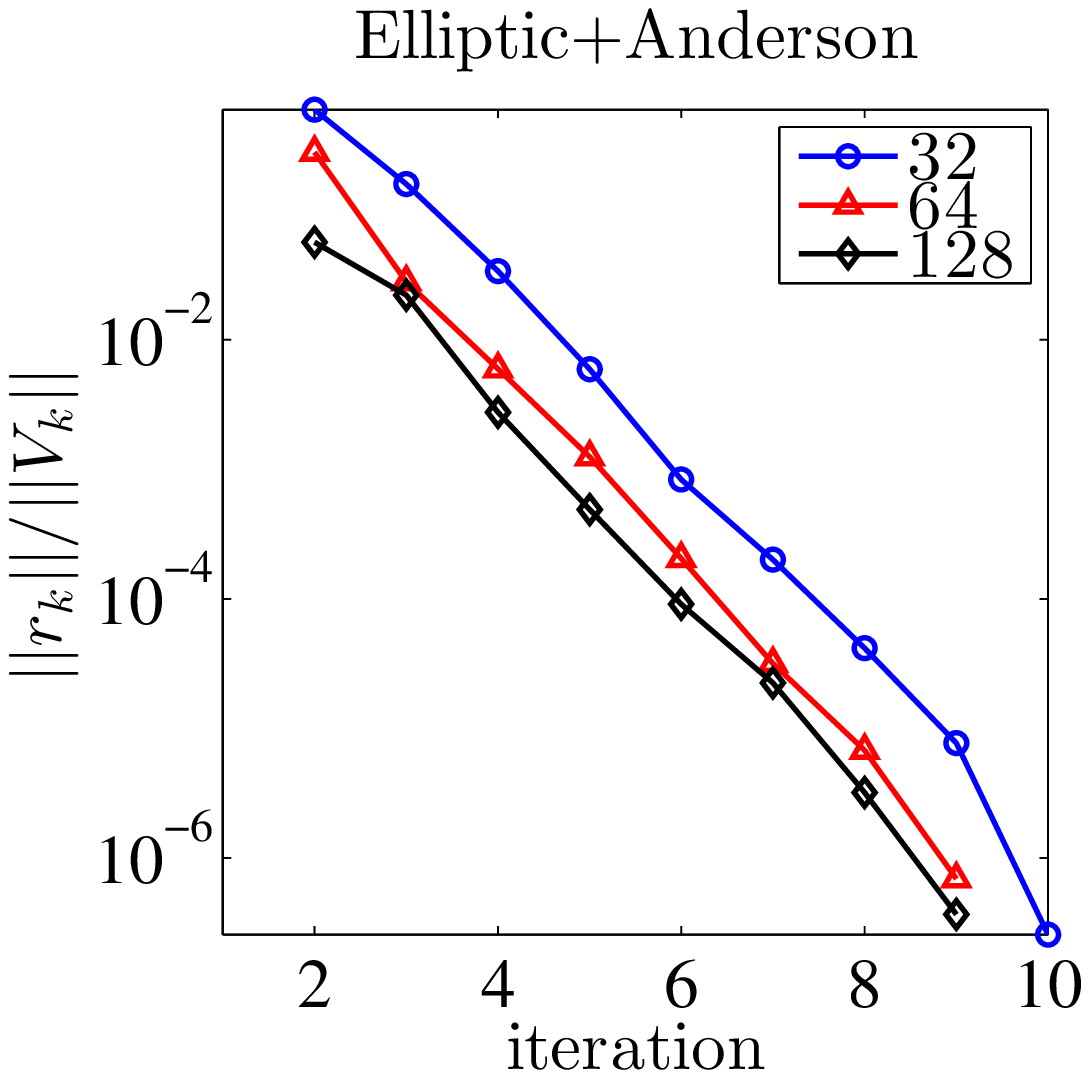}}\\
    \subfloat[Metal+Insulator]{\includegraphics[width=0.3\textwidth,height=0.3\textwidth]{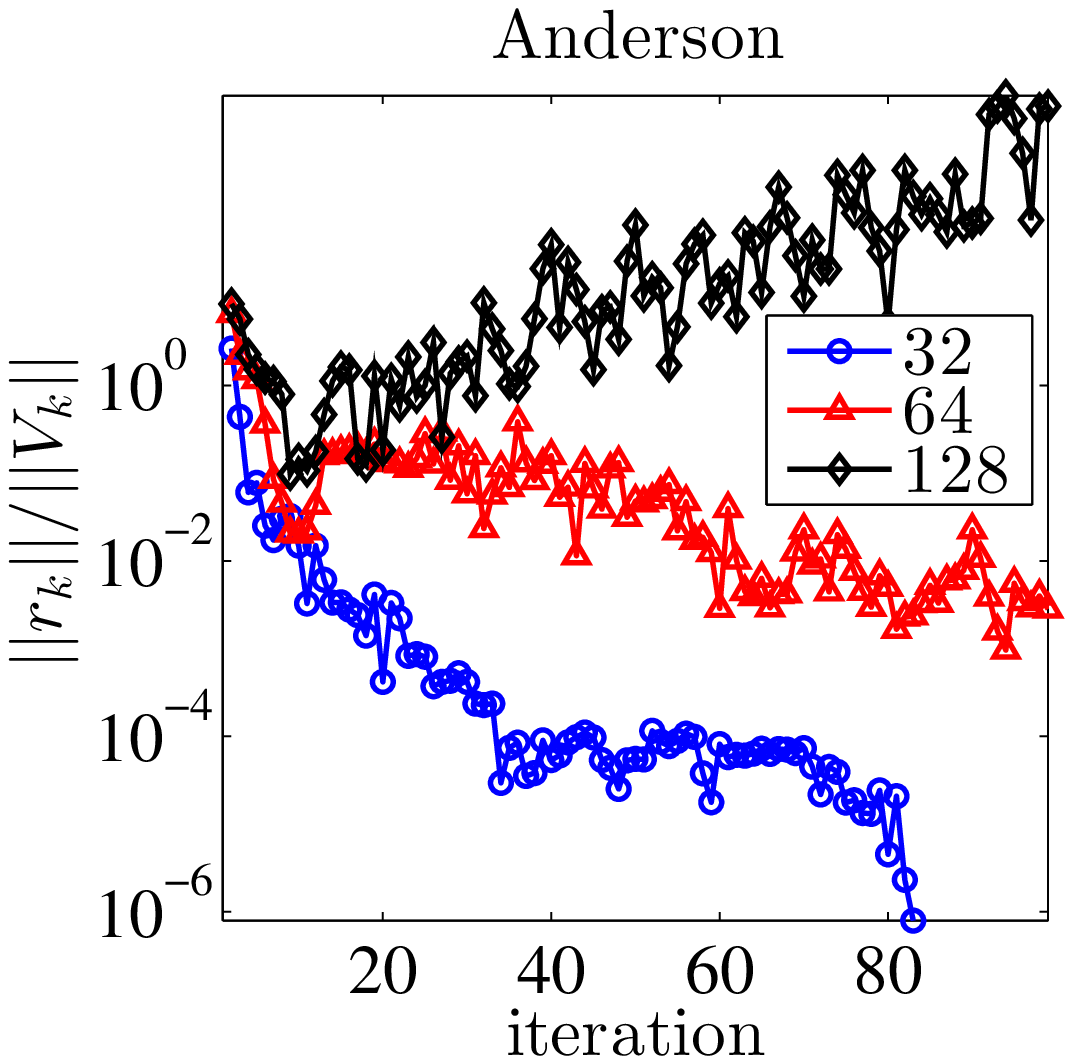}}
    \subfloat[Metal+Insulator]{\includegraphics[width=0.3\textwidth,height=0.3\textwidth]{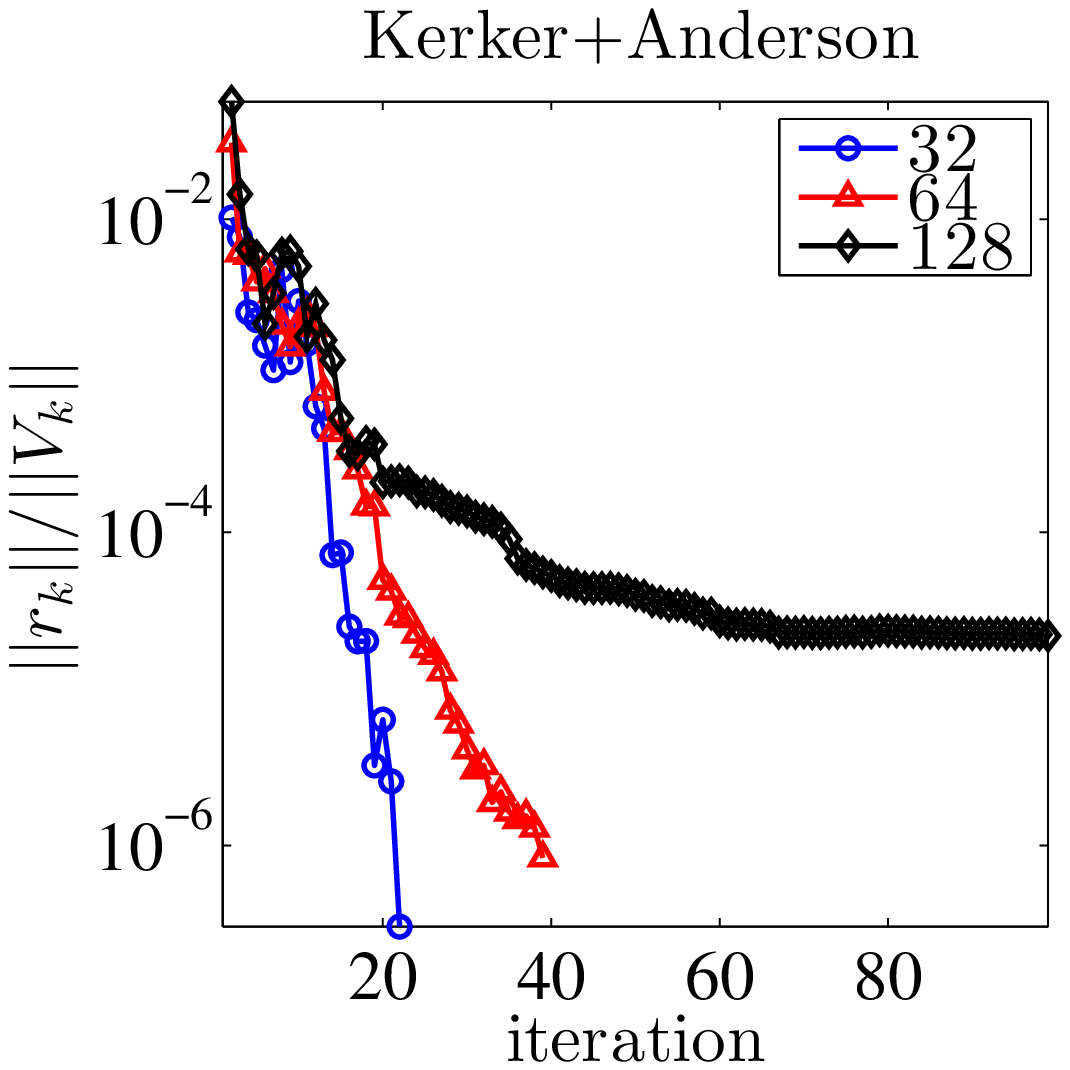}}
    \subfloat[Metal+Insulator]{\includegraphics[width=0.3\textwidth,height=0.3\textwidth]{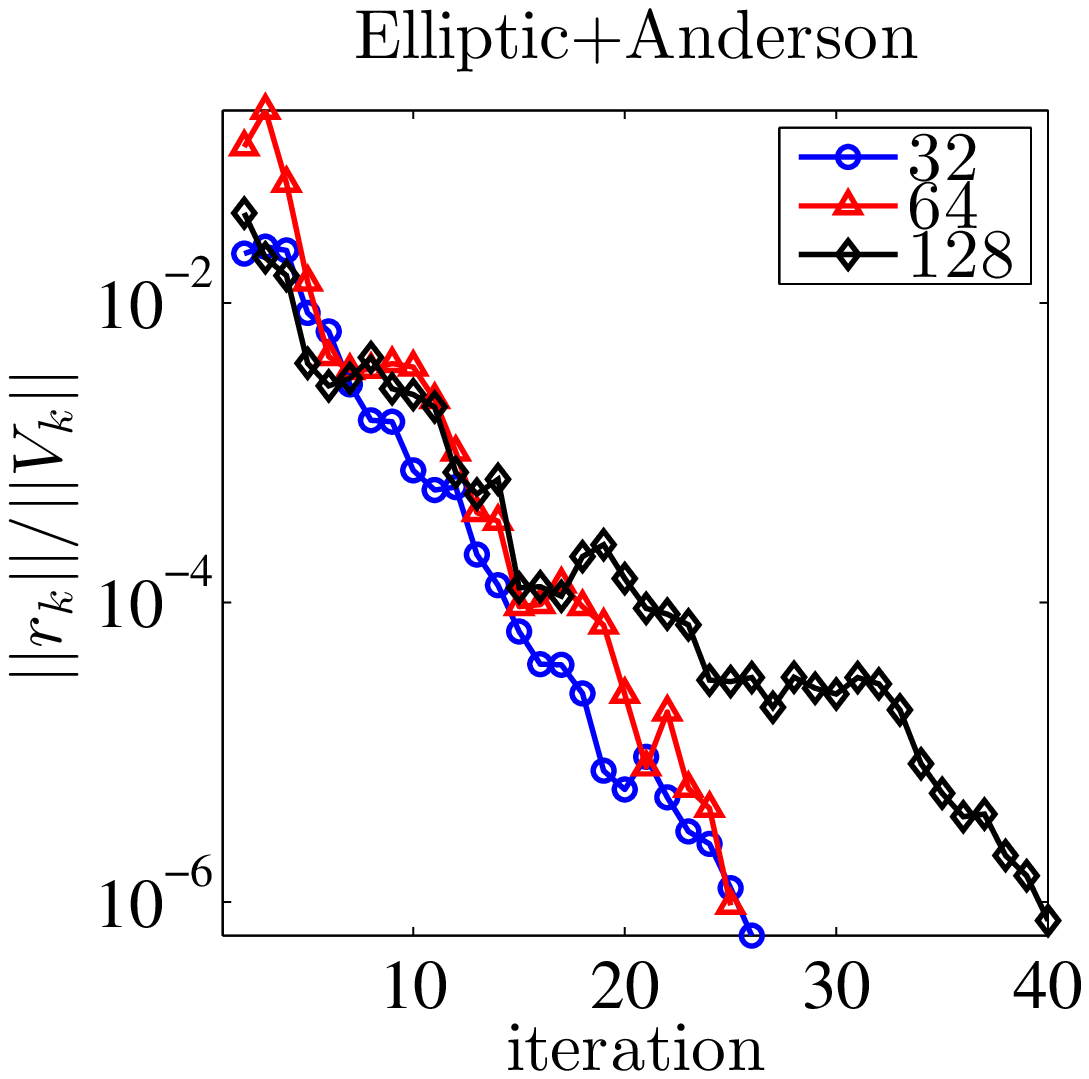}}
  \end{center}
  \caption{The convergence of Anderson's method, Anderson's
  method with the Kerker preconditioner, and Anderson's method with
  the elliptic preconditioner for insulators in subfigures (a), (b), (c),
  for metals in subfigures (d), (e), (f), and for hybrid systems in
  subfigures (g), (h), (i), respectively.}
  \label{fig:converge1d}
\end{figure}

In Figure~\ref{fig:converge1d}, we show the convergence behavior of all
three acceleration schemes for three test cases by plotting the relative
self-consistency error in potential against the iteration number. 
In each one of the subfigures, the blue line with circles, the red line
with triangles and the black line with triangles correspond to 
tests performed on a 32-atom, 64-atom and 128-atom system respectively. 
We observe that the combination of Anderson's method and the elliptic preconditioner
gives the best performance in all test cases. In particular, the 
number of SCF iterations required to reach convergence is more or less 
independent from the type of system and system size.
We can clearly see that the use of the Kerker preconditioner leads to 
deterioration in convergence speed when the system size increases for
insulating and hybrid systems.
On the other hand, Anderson's method alone is not sufficient 
to guarantee the convergence of SCF iteration for metallic and hybrid systems.
All these observed behaviors are consistent with the analysis we 
presented in the previous section.

%

\subsection{Three dimensional sodium system with vacuum}

In this subsection, we compare the performance of different preconditioning
techniques discussed in section~\ref{sec:advance} when they are applied to a 
3D problem constructed in KSSOLV~\cite{YangMezaLeeEtAl2009}, 
a MATLAB toolbox for solving Kohn-Sham problems for molecules and solids.  
We have chosen to use the KSSOLV toolbox because of its ease of use, 
especially for prototyping new algorithms. The results presented here can 
be reproduced by other more advanced DFT software packages such as 
Quantum ESPRESSO~\cite{GiannozziBaroniBoniniEtAl2009}, with some 
additional programming effort.

\begin{figure}[h]
  \begin{center}
    {\includegraphics[width=0.8\textwidth]{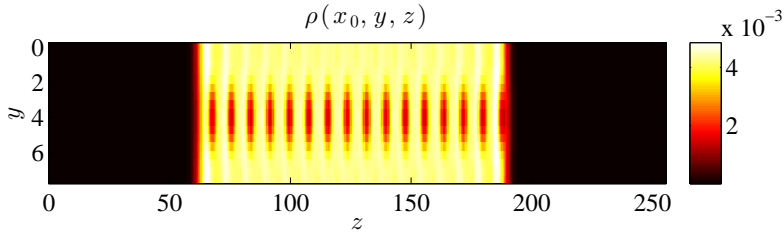}}\\
  \end{center}
  \caption{The $x=0$ slice of the electron density $\rho(x,y,z)$ 
           of the 32-atom sodium system with large vacuum regions at both
           ends.  }
  \label{fig:NaConfig}
\end{figure}

The model problem we construct consists of a chain of sodium atoms
placed in a vacuum region that extends on both ends of the chain.
The sodium chain contains a number of body-centered cubic (BCC) unit cells.
The dimension of the unit cell along each direction is $8.0$ a.u..
Each unit cell contains two sodium atoms.  To examine the 
size dependency of the preconditioning techniques, we tested 
both a 16-unit cell (32-atoms) model and a larger 32-unit cell (64 atoms)
model.  The converged electron density on the $x=0$ plane (or the [100] plane in
crystallography terminology) associated with the 32-atom model
is shown in Figure~\ref{fig:NaConfig}.

\begin{figure}[h]
  \begin{center}
    {\includegraphics[width=0.8\textwidth]{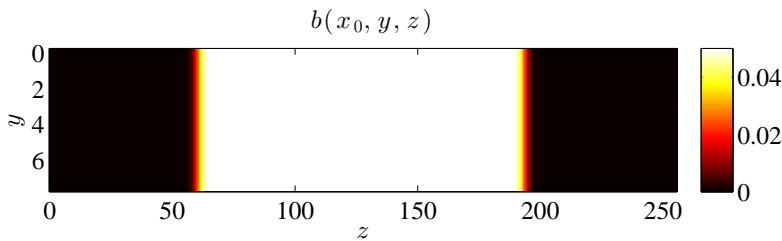}}
  \end{center}
  \caption{The $x=0$ slices of the function $b(x,y,z)$ used in 
  the elliptic preconditioner for a 32-atom sodium system with 
  large vacuum regions at both ends.}
  \label{fig:NaEllipticB}
\end{figure}

\begin{figure}[h]
  \begin{center}
    \includegraphics[width=0.35\textwidth]{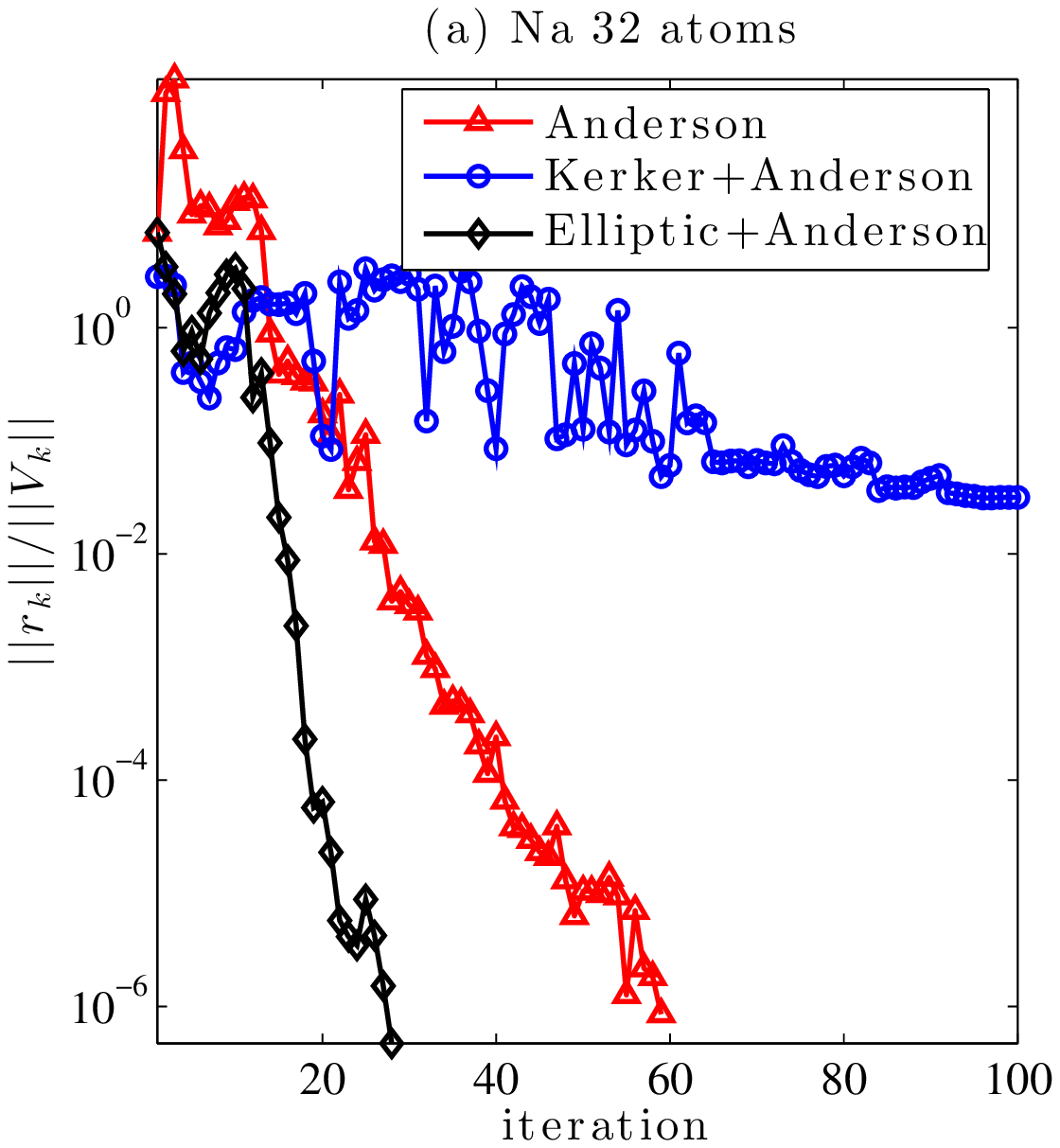}
    \qquad
    \includegraphics[width=0.35\textwidth]{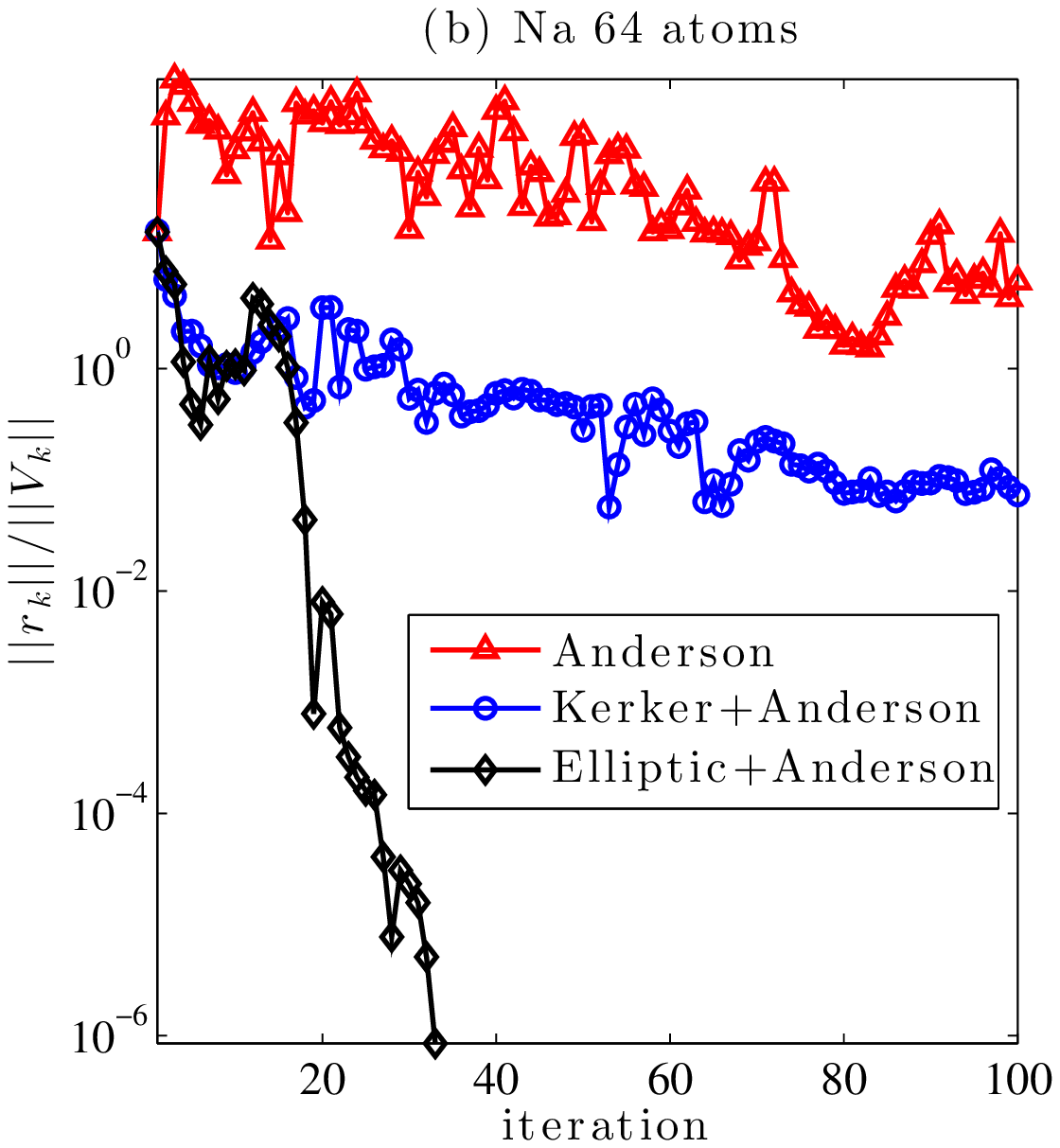}
  \end{center}
  \caption{The convergence of Anderson's method, Anderson's
  method with the Kerker preconditioner, and Anderson's method with the elliptic
  preconditioner for quasi-1D Na systems with a large vacuum region with
  $32$ Na atoms (a) and $64$ Na atoms (b).}
  \label{fig:convNa}
\end{figure}

Figure~\ref{fig:convNa} shows how Anderson's method, 
the combination of Anderson's method and the Kerker preconditioner and 
the combination of Anderson's method and the elliptic preconditioner
behave for both the 32-atom and the 64-atom sodium systems. 
For the 32-atom problem, the parameter $\alpha$ for the 
Anderson's method is set to $0.4$. The parameter $\hat{\gamma}$ required 
in both the Kerker preconditioner and the elliptic preconditioner is set to 0.05.
For the 64-atom problem, the parameter $\alpha$ is set to 
$0.8$. For simplicity the function $a(x)$ required in the elliptic
preconditioner is set to a constant function $a(x) = 1.0$. The $b(x)$
function (shown in Figure~\ref{fig:NaEllipticB} for the 32-atom problem)
is constructed by convolving a square wave function with a value of
$0.05$ in the sodium region and $0$ in the vacuum region with a Gaussian
kernel.  The SCF iteration is declared to be converged when the relative
self-consistency error in the potential is less than $10^{-6}$.

As we can clearly see from Figure~\ref{fig:convNa}, the use of the
Anderson's method with the elliptic preconditioner leads to rapid convergence.
Furthermore, the number of iterations (around 30) required to reach 
convergence does not change significantly as we move from the 32-atom 
problem to the 64-atom problem.

Using Anderson's method alone enables us to reach convergence 
in 60 iterations for the 32-atom problem. However, it fails to 
reach convergence within 100 iterations for the 64-atom case.  
When Anderson's method is combined with the Kerker preconditioner,
the SCF iteration converges very slowly for both the 32-atom and the
64-atom problems.

\section{Concluding Remarks}\label{sec:conclusion}
We discussed techniques for accelerating the convergence of 
the self-consistent iteration for solving the Kohn-Sham problem.
These techniques make use of the spectral properties of the Jacobian 
operator associated with the Kohn-Sham fixed point map.  
They can also be viewed as preconditioners 
for a fixed point iteration.  We pointed out the crucial difference 
between insulating and metallic systems and different strategies 
for constructing preconditioners for these two types of 
systems.  A desirable property of the preconditioner is that
the number of fixed point iterations is independent of the size of 
the system. We showed how this property can be maintained
for both insulators and metals. Furthermore, we proposed a new 
preconditioner that treats insulating and metallic systems in a unified way. 
This preconditioner, which we refer to as an elliptic 
preconditioner, is constructed by solving an elliptic PDE with 
spatially dependent variable coefficients.  Constructing 
preconditioners for insulating and metallic systems simply amounts
to setting these coefficients to appropriate functions.
The real advantage of this type of preconditioner is that it allows
us to tackle more difficult problems that contain both 
insulating and metallic components at low temperature.  We showed by
simple numerical examples that this is indeed the case.  
Although the size of the systems used in our examples are relatively
small because we are limited by the use of MATLAB, we can already
see the benefit of an elliptic preconditioner in terms of keeping
the number of SCF iterations relatively constant even as the system
size gets larger. To fully test whether the preconditioner 
can achieve the goal of keeping the SCF iterations system size
independent, we should implement the elliptic 
preconditioner in a standard electronic structure calculation software 
packages such as QUANTUM ESPRESSO~\cite{GiannozziBaroniBoniniEtAl2009}, 
ABINIT~\cite{Abinit1} and SIESTA~\cite{SolerArtachoGaleEtAl2002} etc. 
that are properly parallelized, which we plan to do in the near 
future.

\section*{Acknowledgments}

This work was supported by the Laboratory Directed Research and
Development Program of Lawrence Berkeley National Laboratory under
the U.S. Department of Energy contract number DE-AC02-05CH11231 (L. L. 
and C. Y.).  We would like to thank Eric Canc\`{e}s, Roberto Car, Weinan E,
Weiguo Gao, Jianfeng Lu, Lin-Wang Wang and Lexing Ying for helpful
discussion. 



\end{document}